\documentclass[apj]{emulateapj}
\usepackage{apjfonts}
\usepackage{lscape}
\def\teff{T$_{\rm eff}$}
\def\logg{log $g$}

\begin{document}

\title{Detailed Abundances for 28 Metal-poor Stars: Stellar Relics
  in the Milky Way\altaffilmark{1,2}}

\author{David K. Lai{\altaffilmark{3}}, Michael Bolte{\altaffilmark{3}}, Jennifer
  A. Johnson{\altaffilmark{4}}, Sara Lucatello{\altaffilmark{5,6}}, Alexander
Heger{\altaffilmark{3,7}}, and S. E. Woosley{\altaffilmark{3}}}
\altaffiltext{1}{The data presented herein were obtained at the W.M. Keck Observatory, which is operated as a scientific partnership among the California Institute of Technology, the University of California and the National Aeronautics and Space Administration. The Observatory was made possible by the generous financial support of the W.M. Keck Foundation.} 
\altaffiltext{2}{This publication makes use of data products from the Two Micron All Sky Survey, which is a joint project of the University of Massachusetts and the Infrared Processing and Analysis Center/California Institute of Technology, funded by the National Aeronautics and Space Administration and the National Science Foundation.}
\altaffiltext{3}{Department of Astronomy and Astrophysics, University
  of California, Santa Cruz, CA 95064; david@ucolick.org,
  bolte@ucolick.org, alex@ucolick.org, woosley@ucolick.org.}
\altaffiltext{4}{Department of Astronomy, Ohio State University, 140
  W. 18th Ave., Columbus, OH 43210; jaj@astronomy.ohio-state.edu.}
\altaffiltext{5}{Osservatorio Astronomico di Padova, Vicolo dell'Osservatorio 5, 35122 Padua, Italy; sara.lucatello@oapd.inaf.it.}
\altaffiltext{6}{Excellence Cluster Universe, Technische Universit\"{a}t
  M\"{u}nchen, D-85748 Garching, Germany.}
\altaffiltext{7}{Theoretical Astrophysics Group, T-6, MS B227, Los Alamos National Laboratory, Los Alamos, NM 87545.}

\begin{abstract}
We present the results of an abundance analysis for a sample of stars
with $-4<$[Fe/H]$<-2$. The data were obtained with the HIRES
spectrograph at Keck Observatory. The set includes 28 stars, with
effective temperature ranging from 4800 to 6600 K. For 13 stars with
[Fe/H]$<-2.6$, including nine with [Fe/H]$<-3.0$, and one with
[Fe/H]$=-4.0$, these are the first reported detailed abundances. For
the most metal-poor star in our sample, CS 30336-049, we measure an
abundance pattern that is very similar to stars in the
range [Fe/H]$\sim-3.5$, including a normal C+N abundance. We also find
that it has very low but measurable Sr and Ba, indicating some
neutron-capture activity even at this low of a metallicity. We explore
this issue further by examining other very neutron-capture-deficient
stars, and find that at the lowest levels, [Ba/Sr] exhibits the ratio
of the main $r$-process. We also report on a new $r$-process-enhanced
star, CS 31078-018. This star has [Fe/H]$=-2.85$, [Eu/Fe]$=1.23$,
and [Ba/Eu]$=-0.51$. CS 31078-018 exhibits an ``actinide boost'',
i.e. much higher [Th/Eu] than expected and at a similar level to
CS 31082-001. Our spectra allow us to further constrain the abundance
scatter at low metallicities, which we then use to fit to the zero-metallicity
Type II supernova yields of \citet{heger08}. We find that supernovae with progenitor masses between 10
and 20 M$_{\odot}$ provide the best matches to our abundances.

\end{abstract}


\section{Introduction}
In recent years the number of discovered extremely metal-poor (EMP,
[Fe/H]$\leq-3.0$) star candidates has grown substantially, thanks in
large part to the survey of \citet{bps}, and the more recent
Hamburg/ESO (HES) survey \citep{christlieb00}. The high-resolution
follow-ups to these surveys (e.g. \citealt{mcwilliam95};
\citealt{cohen04,cohen07-2}; \citealt{cayrel04}; \citealt{aoki05}; \citealt{heres2}) have verified about 100 stars with [Fe/H]$<-3.0$.

These EMP stars play an important role in understanding the very first
generation of stars (\citealt{beers05} and references and discussion
therein). The lower the metal content of a star, the fewer instances
of nucleosynthesis and recycling that preceded its formation. For the
most metal-poor stars we may have the opportunity to measure the
undiluted imprint of Population III nucleosynthesis. The best examples
of this possibility are the two most metal-poor stars known, HE
1327-2326 \citep{frebel05} and HE 0107-5240 \citep{christlieb02}. Both
stars have [Fe/H]$\sim-5.3$, and their abundance ratios can be fitted by
zero-metallicity supernovae (SNe) with a tuned mixing parameter in the
pre-ejecta material \citep{iwamoto}. An alternative scenario to
explain the HE 0107-5240 abundance ratios is proposed by
\citet{suda04}, in which it is a Population III star that has accreted
its heavy elements through binary and interstellar medium (ISM) accretion. In either case we
are most likely seeing the imprint of the first stars, whether it is
in SN ejecta or asymptotic giant branch (AGB) evolved material. However, there is a third
possibility put forth by \citet{venn08}, that these stars are of a
class of chemically peculiar stars that have true [Fe/H] values
greater than $-4.0$.

As the sample of EMP stars has grown, a curious feature of the
metal-poor end of the metallicity distribution function (MDF) of stars
in the Galaxy has become apparent. Recently, \citet{norris07} discovered a
star with [Fe/H]$=-4.8$, HE 0557-4840, making a total of only three
stars with [Fe/H]$<-4.1$. However, when including these stars in
the MDF, as \citet{norris07} point out, the number of stars
discovered with metallicities $-5.3<$[Fe/H]$<-4.1$ still falls 3-4
times short of what is expected from the mixing and fallback models of
chemical enrichment (e.g., \citealt{salvadori}). Instead, a two-component
model, as described by \citet{karlsson}, where feedback
effects from the first massive stars inhibit more star formation, may
better fit the statistics of the observed halo MDF. 

One key to understanding these issues is to increase the sample of
well-studied EMP stars to the point where the different classes of
abundance patterns can be identified and therefore explore the nature
of their progenitors. While most EMP stars with [Fe/H]$>-3.5$ have relatively small dispersions for elements at or below the iron-peak
(e.g. \citealt{carretta02} and \citealt{cayrel04}), recent observations have
shown that there are some objects that are either strongly enhanced or
deficient in certain $\alpha$-elements (\citealt{aoki07b} and
references therein). At the more metal-poor end, \citet{cohen07}
report a star with [Fe/H]$\sim-4.0$ and a highly unusual
abundance pattern, HE 1424-0241. It has very low [Ca/Fe] and [Si/Fe],
$-$0.58 and $-$1.01, respectively, but has a [Mg/Fe] of 0.44,
which is typical of EMP stars. \citet{cohen07} find that there are
no core-collapse SN models that can fit this abundance pattern. These
important results re-emphasize the need to find more stars in this
metallicity regime. Only then can we begin to find out what, if
anything, is typical, and it is clear with HE 1424-0241 that we must be
very careful extending trends from [Fe/H]$>-3.5$ to lower
metallicities.

EMP stars also exhibit a wide dispersion in the ratio of the
neutron-capture elements to iron
(e.g. \citealt{mcwilliam98,honda04,francois}). By examining both
the most neutron-capture rich and neutron capture-poor EMP stars we
can shed light on the different processes that give rise to this
dispersion. Beginning with the discovery of CS 22892-052
\citep{sneden96,sneden03}, the imprint of a universal $r$-process
pattern has been found to stretch from EMP stars up to the Sun. As
more $r$-process stars are discovered, this result has been even more
strongly confirmed (at least for elements with Z$\geq56$). For the
$s$-process we are also beginning to see a convergence between
observational abundance ratios of the neutron-capture elements in
these stars and models of EMP AGB stars
(e.g. \citealt{jbcs22183,jbcs31062, masseron06}). Even though some
tuning of the $^{13}$C pocket formation in the AGB star is needed, the
match between observed abundances and models is encouraging. These
agreements, however, for both the $r$-process and $s$-process, do not explain the
origin of some of the lighter neutron-capture elements (Z$<56$). To
investigate this, the measurement of the light neutron-capture element
strontium may prove an ideal probe, as its resonance lines are still
detectable in EMP stars. Part of the answer to this puzzle may come
from looking at the most neutron-capture poor stars, to isolate the
other process(es) that contribute to these elements.

In this study we present abundance ratios from [C/Fe] to [Eu/Fe] for
stars in various evolutionary states in the metallicity range
$-4<$[Fe/H]$<-2$. In addition to the various nucleosynthesis events
described above, mixing that occurs as a star evolves from the main
sequence and up the giant branch can also affect the light element
abundances up to N \citep{gratton00,spite05,spite07}. Our data
allows us to both see these evolutionary effects and provide a
picture of the early Galaxy through its nucleosynthetic footprint. They
also reveal unexpected correlations of \teff{} with \ion{Si}{1}, 
\ion{Ti}{1} and \ion{Ti}{2}, and \ion{Cr}{1}.

In $\S$ 2 and $\S$ 3 we present the details of the observations and
analysis. In $\S$ 4 we present the abundance results of our study and
compare them to previous samples of metal-poor stars. In $\S$ 5 we
discuss interesting individual stars, as well as properties of the
sample as a whole, including the curious behavior of \ion{Si}{1},
\ion{Ti}{1}, \ion{Ti}{2}, and \ion{Cr}{1}. Also in this section we
present fits to the zero-metallicity Type II SNe (Sne II)
from \citet{heger08}.

\section{Observations and Reductions}
We chose our sample from \citet{lai04} and
metal-poor candidates identified in the photometric sample of \citet{schuster04}.
The data were obtained from multiple runs at the HIRES spectrograph
\citep{vogt94} at Keck Observatory between 2001 and 2006. A detector
upgrade in mid-2004 allowed us to obtain higher quality spectra in the
blue region.  Before the upgrade we typically observed using a blue
and red configuration for HIRES, and after the upgrade we observed with
a single setup.  The details of the observations, including wavelength
coverage, signal-to-noise ratio (S/N), and $V$ magnitude, are given in Table \ref{odetails}.

The reductions were done differently for the data before and after the
detector upgrade. The pre-upgrade spectra were reduced with the MAKEE (Mauna
Kea Echelle Extraction) reduction package.  The post-upgrade data were
reduced with the HIRES data reduction package written by
J. X. Prochaska.

\subsection{Equivalent Widths \label{ewsection}}

We used the spectrum analysis code SPECTRE \citep{spectre} to measure
individual equivalent widths (EWs) of isolated lines. The bulk of
lines were measured by Gaussian fitting, but for some of the stronger
lines we used a Voigt profile to better fit the line wings.

\begin{figure}
\begin{center}
\scalebox{.45}[.45]{
\plotone{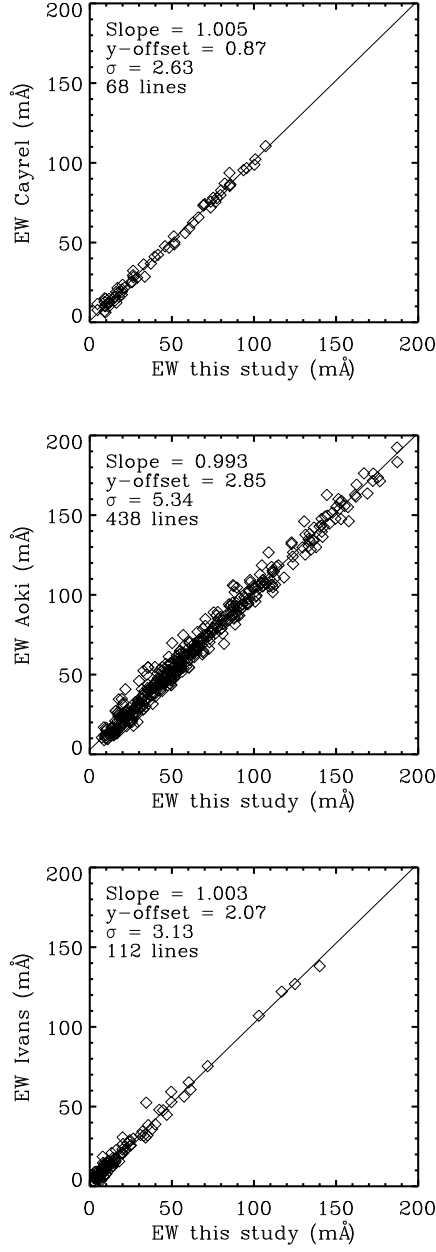}}
\end{center}
\figcaption{The top panel compares the EWs measured by
  \citet{cayrel04} with our study, the middle panel shows the
  comparison with \citet{aoki05}, and the bottom panel shows the
  comparison with \citet{ivans03}.  \label{ewcomp}}
\end{figure}

As a check of our accuracy and data quality, we compare our EW
measurements with previous high-resolution studies done of common
objects. Figure \ref{ewcomp} compares the EWs from the studies of
\citet{cayrel04}, \citet{aoki05}, and \citet{ivans03} to our
measurements. We share 63 common lines with the Cayrel study for the
star BS 16467-062, 438 common lines with the Aoki study from
the four stars BS 16080-054, BS 16084-160, CS 30312-059, and CS 30325-028,
and 112 common lines with the Ivans study for the stars BD+03 740
and BD+24 1676. As can be seen from the figure, our values are in very
good agreement with the three studies. On average we are finding
slightly lower EWs than these previous studies of 1.09, 
2.41, and 2.24 m\AA{}, as compared to \citet{cayrel04}, \citet{aoki05},
and \citet{ivans03}, respectively. 

\section{Stellar Parameters and Analysis}

We used a current version of Turbospectrum \citep{turbo}, which
properly accounts for continuum scattering (see \citealt{cayrel04}), in
combination with the stellar atmospheres from \citet{castelli03} to perform LTE line analysis and spectral synthesis. Our atomic
line data along with measured EWs are given in Tables
\ref{ew1} through \ref{ew3}. We began with the line lists from
\citet{ivans06} and \citet{sneden03}, and added additional lines found
using the NIST atomic line database. Specifically we updated/added the
following $gf$-values: \ion{Fe}{1} and \ion{Fe}{2} lines with those of \citet{felines}, the Mg
lines at 5172.7 and 5183.6 \AA{} with the values from
\citet{mgtriplet}, \ion{Cr}{1} lines with values from \citet{sobeck07}, \ion{Cr}{2}
lines with values from \citet{crIIlines}, the 
\ion{Mn}{1} lines at 3577.9 and 4055.6 \AA{} with values from
\citet{mnIupdate}, and \ion{Zr}{2} lines with values from \citet{ljung},
when available, or otherwise from \citet{malcheva}, when available.

To measure oxygen, we adopted the linelist from Kurucz\footnote{http://kurucz.harvard.edu} for the OH
region at 3185 \AA{} with a dissociation potential of 4.40
eV. We measured nitrogen from the NH feature at 3360 \AA{}. Following
the prescription from \citet{johnson07} for this list, we used the
Kurucz $gf$-values divided by 2 with a dissociation potential of 3.45
eV. The CH linelist at 4300 \AA{} was derived from the LIFBASE
database (courtesy of B. Plez).

\subsection{Hyperfine splitting}

The effects of hyperfine splitting (hfs) can greatly
affect the derived abundance from strong lines of certain
elements. The effect is a desaturation of strong lines, and therefore
a larger equivalent width than would be found given the absence of the
splitting. \citet{prochaska00} noted the effects of hfs for Sc and Mn.
We have taken the hfs parameters from Kurucz to account for these two
elements. Vanadium and cobalt are also known to be affected by hfs;
however, all of our lines are below 20 m{\AA}, too weak for it to
change their derived abundances. We have adopted the hfs parameters
and isotope ratios of Ba from \citet{mcwilliam98} along with updated $gf$-values used by
\citet{ivans06}. The hfs parameters for Eu are taken from
\citet{sneden03}. The near UV-lines that we use to measure Cu are also known to be
greatly affected by hfs \citep{bihain}. We account for this by using the hfs
parameters from Kurucz and assuming the solar isotope ratio of $^{63}$Cu
to $^{65}$Cu from \citet{anders89}.

\subsection{Radial Velocities}

Radial velocities were determined by cross-correlating our program
star spectra with high-S/N template stars using FXCOR.\footnote{IRAF
is distributed by the National Optical Astronomy Observatories, which
are operated by the Association of Universities for Research in
Astronomy, Inc., under cooperative agreement with the National Science
Foundation.} These template star spectra were taken during the same
observing run as the respective program stars, and their radial
velocities were measured with individual strong absorption lines
(typically 50 lines). These results are given in Table
\ref{odetails}. The typical internal error from this procedure is only
$\sim$0.2 km s$^{-1}$. However this does not take into account the
overall zero-point error, particularly given the diversity of our
instrument setups. A more realistic estimate to the error in absolute
radial velocity can be had by looking at velocities derived for the
same object observed during different nights with different instrument
setups, but on the same run (i.e. 2004 April 7$-$9). We find that there
is an rms of $\sim$1.0 km s$^{-1}$ in these measurements. In our sample
BS 16084-160 is clearly identified as a radial velocity variable. Also
BS 16550-087 is a likely radial velocity variable.

\subsection{Model Atmosphere Parameters}

\subsubsection{\teff{}}
We obtained our effective temperatures using the $V-K$ color of each
star.  The $K$ magnitudes were taken from the Two Micron All Sky
Survey. We then transformed the color using the updated \citet{alonso96,alonso99} color-\teff{} calibration given by
\citet{ramirez05}. The reddenings were taken from the \citet{schlegel} dust maps, except for values of E(B-V) greater than 0.10.
The Schlegel map may overestimate reddening for higher values \citep{arce99};
we adopted Equation 1 from \citet{bonifacio00} to account for
this.

\begin{figure}
\begin{center}
\scalebox{.8}[.8]{
\plotone{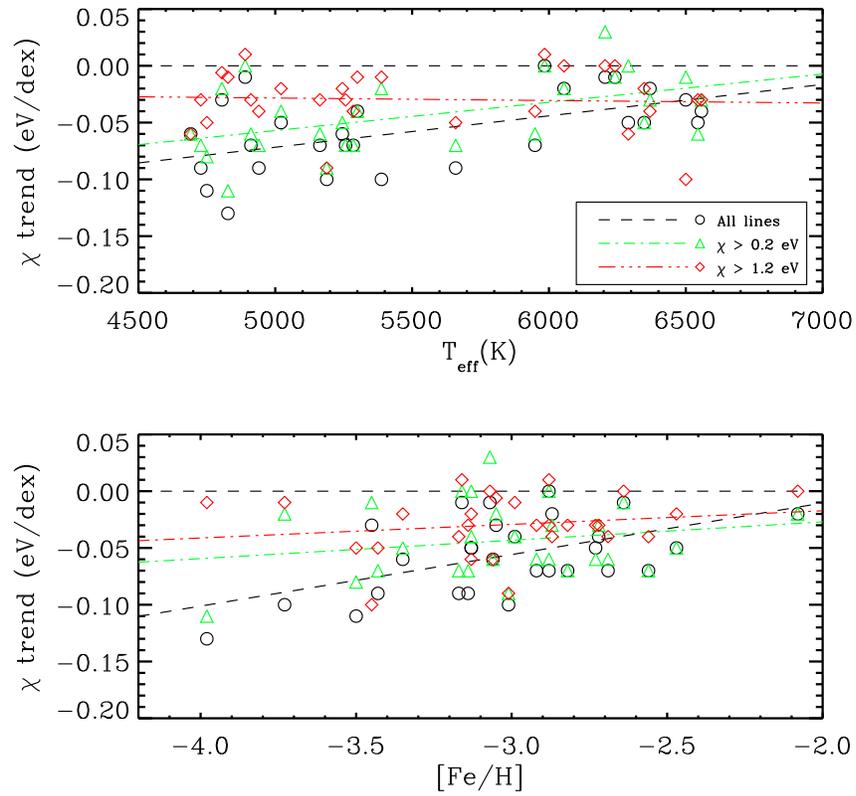}}
\end{center}
\figcaption{Value of the $\chi$/log$\epsilon$(Fe {\scriptsize I})
  slopes for each of our stars. When all Fe I lines are considered,
  there seems to be a trend with both \teff{} and [Fe/H], although
  considering only the $\chi>$1.2 eV lines, this trend largely
  disappears. Overall, however, even when considering high $\chi$
  lines, we still seem to be finding a negative slope in most of our
  stars. See the electronic edition of the Journal for a color
  version of this figure. \label{eptrend}}
\end{figure}

We find a trend in \ion{Fe}{1} lines with excitation potential
($\chi$) in many of our stars. This trend is usually accounted for by
adjusting the \teff{}. In Table \ref{atm} we give the value for the
slope of the trends, the corresponding correlation coefficient ($r$),
and the number of \ion{Fe}{1} lines used, for different cuts on the minimum
$\chi$ considered. Almost all of our stars have a negative slope when considering
all \ion{Fe}{1} lines, which implies that our \teff{} is too high. Unlike
\citet{cohen07-2}, we find that using the 0.2 eV cut still leaves a
statistically significant correlation between individual \ion{Fe}{1} lines
and their corresponding $\chi$-values. \citet{cayrel04} found that a cut at
1.2 eV seemed to eliminate the trends they found, and it does seem to
markedly reduce the trends and their statistical significance in this study. However,
unlike in \citet{cayrel04}, small but marginally significant trends
still exist. In Figure \ref{eptrend} we show this effect for the above
cuts on $\chi$.

To test the accuracy of our \teff{} determinations, we also fit
the Balmer lines of two of our stars, CS 22880-086 and CS
30336-049. CS 22880-086 in particular shows very little reduction in
the trends with the $\chi$ cuts, and CS 30336-049 has the largest trend
with $\chi$ when considering all \ion{Fe}{1} lines. In neither star do we
observe H$\alpha$, and the H$\beta$ lines are positioned on the edges
of the echelle orders. We therefore fit H$\gamma$ and H$\delta$. For
both stars these two Balmer lines are fitted quite well with the $V-K$-derived \teff{}. Part of the answer to this discrepancy between
spectroscopic temperature with color and Balmer line temperatures may come from
the three-dimensional (3D) hydrodynamical effects in these
stars. \citet{collet} recently examined the potential impact of this
in red giant stars. For metal-poor stars they find a strong effect
that would explain the direction of our trend but overestimates the
magnitude. For this study we have chosen to stay with the $V-K$-derived
temperatures and note that the effects of inaccurate \teff{} are
minimized by looking at abundance ratios versus absolute abundances.

\subsubsection{Surface gravity and Microturbulent velocity}

The \teff{} was then used to determine the surface gravity.  We used
the Y$^2$ isochrones from \citet{kim2002}, with $\alpha$-enhancement set
to 0.3 and age to 12 Gyr.  We chose not to adjust the \logg{}
spectroscopically to get the abundance of iron from \ion{Fe}{2} and
\ion{Fe}{1} to match.  This way we avoid the potential non-LTE (NLTE) effects on
the \ion{Fe}{1} abundance giving us an erroneous \logg{}. Even without
adjustment our \ion{Fe}{1} abundances are in generally very good
agreement with the \ion{Fe}{2} abundances.

\begin{figure}
\begin{center}
\scalebox{.7}[.7]{
\plotone{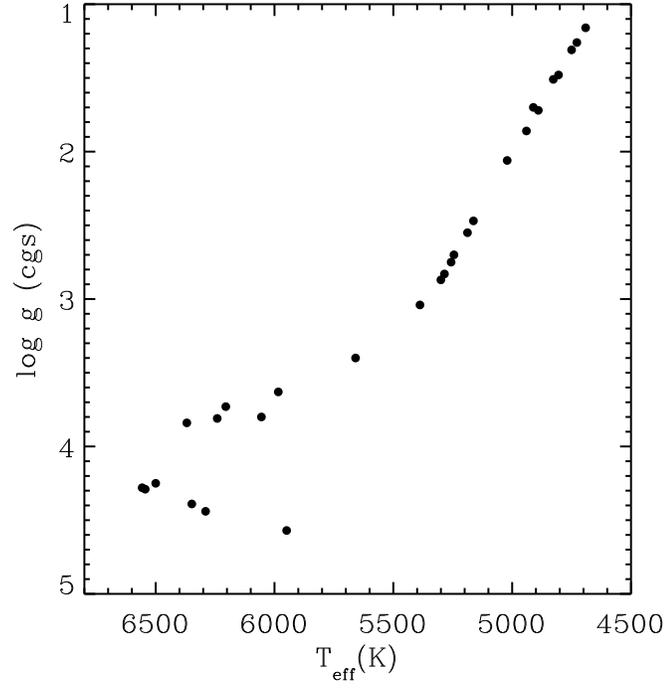}}
\end{center}
\figcaption{Plot of \teff{} vs. \logg{} for our stars. The sample spans a wide range of evolutionary states. \label{tvslogg}}
\end{figure}

The final microturbulent velocity was determined spectroscopically by
eliminating any trend with EW versus abundance for the
\ion{Fe}{1} lines. Because the \teff{}-color calibration is dependent on
metallicity, and therefore the \logg{} as well, we iterated the above
method until we settled on a metallicity within 0.10 dex of our \ion{Fe}{1}
abundance.  The final atmospheric parameters are listed in Table \ref{atm}.
Figure \ref{tvslogg} shows the span of evolutionary states of our sample.

\subsection{Error Analysis}

Excluding systematic NLTE and 1D versus 3D atmosphere effects, the
uncertainties in our measurements come from three sources.  The first
comes from the error in the EW measurement (or in the case of
synthesis, the error in the fit). The second comes from errors in the
atomic parameters. In most cases we have multiple lines measured for
the same element in any given star and the scatter in those lines can
give an estimate for the first two error sources. When four or fewer
lines of an element in a star are measured, we calculate the average
dispersion for the sample in that element, and use this to set a
minimum value. We then adopted the larger of the two values, the
actual dispersion or this minimum value, to estimate this error
component. If there is only a single line measured for a given
element, then we assume an error of 0.15 in its abundance. In the
case of the synthesis, we estimate this component of the error by how
well fitted we can match the synthesis to the actual spectrum (this is
typically 0.1).

The third source of error comes from the uncertainty in the model
atmosphere parameters. We adopt the errors of 100K and 0.2 dex for
\teff{} and \logg{}, respectively. The error estimate for \teff{} from
using broadband colors has been estimated by multiple authors before
(e.g. \citealt{cohen02}), and 100K is a conservative value. The
\logg{} error is more difficult to estimate. Although we do not use
ionization balance to determine surface gravity, as mentioned above we
still find that \ion{Fe}{1} and \ion{Fe}{2} agree very well in
our stars. Changes to \logg{} of 0.2 dex generally generate a
noticeable difference between \ion{Fe}{1} and \ion{Fe}{2}, and we
use this as an error estimate of the isochrones themselves. We estimate the error
for microturbulence velocity to be 0.2 km s$^{-1}$, as at differences greater
than this pronounced trends of equivalent width versus \ion{Fe}{1}
line abundances appeared.

We have adopted the error analysis technique described by
\citet{mcwilliam95} and \citet{johnson2002}. In particular, we use
equations 3, 5, and 6 from \citet{johnson2002}, including the
covariance terms to take into account the dependent nature of our \teff{},
\logg{}, and microturbulence values. Because our sample spans a
wide range of evolutionary states, we use three different stars to
estimate the atmospheric effects, BD+03 740, CS 31078-018, and CS
29502-092, to cover the main-sequence/turnoff, sub- lower giant
branch, and the upper giant branch, respectively. In Tables
\ref{erBD+03}, \ref{erCS31078}, and \ref{erCS29502} we summarize the
results for these three situations. Using equation 5 and 6 from
\citet{johnson2002}, these values can then be used to estimate both the
final absolute and relative errors.

\subsection{Comparison to Previous studies}

As a final check of our method and analysis we compare our atmosphere
and abundance results to those stars from the studies listed in $\S$
\ref{ewsection}. For BS 16467-062, we also consider additional
abundances from \citet{bonifacio07} and the abundance analysis
by \citet{cohen07-2}.

In Table \ref{atmcomp} we summarize the atmospheric parameters. In
general we are in very good agreement with these previous studies. The
agreement with \citet{cohen07-2} is not surprising given that we derive
our parameters in a very similar way. All of the other studies, however,
use ionization balance to estimate surface gravities. That we agree
with these \logg{} values adds confidence to the
isochrone method that we use to derive surface gravity. One exception to this good agreement is
for BD+03 740. \citet{ivans03}, propose two atmospheres for this
star. We have chosen to list in Table \ref{atmcomp} the atmosphere
that most closely resembles ours. \citet{ivans03} go into detail
about various other atmosphere and abundance determinations for this
object. 

\begin{figure}
\begin{center}
\scalebox{.65}[.65]{
\plotone{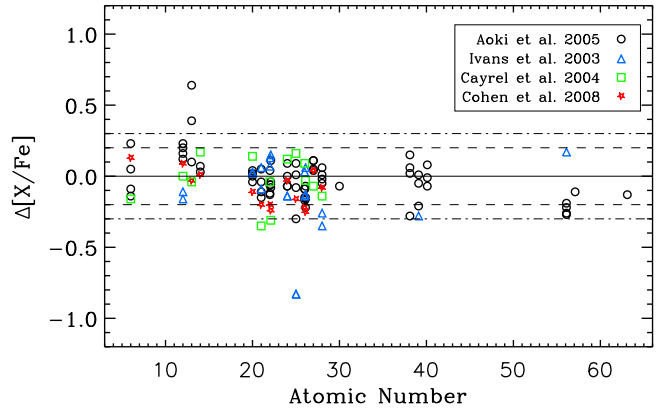}}
\end{center}
\figcaption{Comparison of our abundance to that of other
  studies. The sense of the $y$-axis is the values from this study minus
  the values from previous studies. See the electronic edition of the Journal for a color
  version of this figure.\label{compabund}}
\end{figure}

We show the abundance comparisons with these studies in Figure
\ref{compabund}. In most cases we have good agreement, although there
are a few exceptions. We find the largest discrepancy in [Al/Fe]
compared to \citet{aoki05}. The atomic parameters and atmospheres
agree quite well between this study and those from \citet{aoki05}, so those are not the reason for the discrepancy. We believe
the disagreement may arise from a CH absorption feature that is blended with
the Al $\lambda$3944 line.

Another highly discrepant ratio, this time between our study and
that of \citet{ivans03}, is [Mn/Fe]. This difference comes from
measured EWs. \citet{ivans03} measure only a single \ion{Mn}{1} line at
4823.52 \AA{} to have EWs of 14.5 and 7.3m\AA{} for BD+24 1676 and
BD+03 740, respectively. For the same line we measure EWs of 3.0 and
1.2 m\AA{}. We have both higher S/N and higher resolution spectra than
the \citet{ivans03} study, and for such low EW lines it is conceivable
that their measurements for these two lines were overestimated because
of noise. This may also be the cause for the discrepancy we find in
[Ni/Fe]. For both objects the common lines are very weak, and we find
much lower EWs than \citet{ivans03}.

\section{Results}

In the following section we discuss our abundance results along with those
from various other high-resolution studies. These results are reported
in Tables \ref{abund1} through \ref{abund6}, assuming the solar
abundances from \citet{gs98}. We comment on the specific
elements below, but note that in general the abundances from all of
these studies, including our own, agree remarkably well both in
trends and in scatter. 

\subsection{The light elements: C,N,O, and Li \label{light}}

\begin{figure}
\begin{center}
\scalebox{.4}[.4]{
\plotone{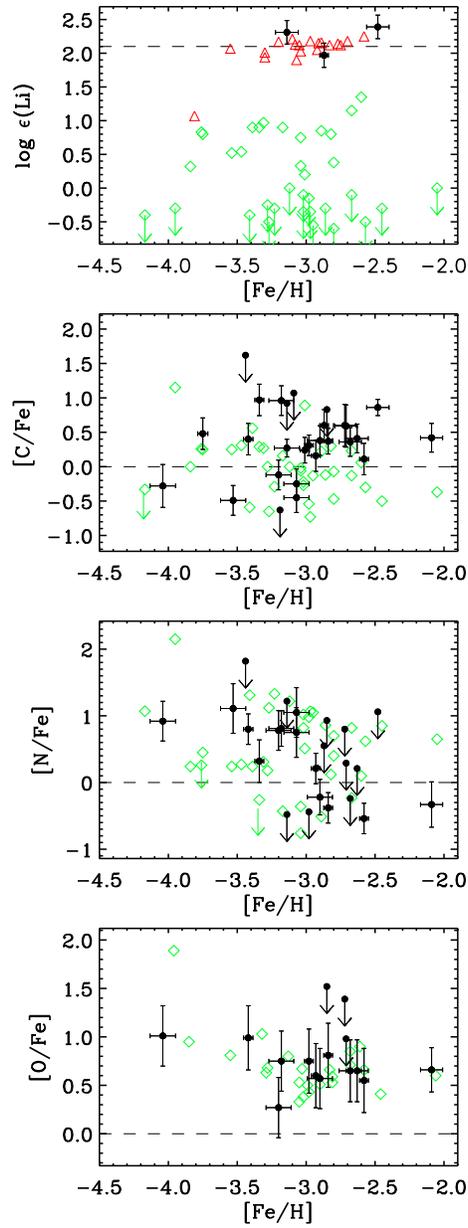}}
\end{center}
\figcaption{Values of Log$\epsilon(\mbox{Li})$ and [(C,N,O)/Fe]
  vs. [Fe/H].  The diamonds (colored green in the electronic edition) are
  from Spite et al. (2005) and the triangles (colored red in the
  electronic edition) are from
  \citet{bonifacio07}. In all of the following figures, the black
  points with error bars are data from this study. See the electronic edition of the Journal for a color
  version of this figure.\label{cno}}
\end{figure}

Figure \ref{cno} shows our values of log$\epsilon(\mbox{Li})$, [C/Fe],
[N/Fe], and [O/Fe] versus [Fe/H].  Overplotted on each are the values
from \citet{spite05}.  For all three elements we find a significant
scatter through the whole range of [Fe/H]. In our most metal-poor
star, CS 30336-049, we find sub-solar [C/Fe].

Although we do find similar values of [O/Fe] as \citet{spite05}, there
should be a word of caution in the manner in which we measure
[O/Fe]. \citet{spite05} use the forbidden \ion{O}{1} line at 6300 \AA{},
while we use the UV OH region at 3185 \AA{}.  According to \citet{asplund01}, taking into account 3D effects may reduce oxygen
abundances derived from the OH lines by as much as 0.6 dex.  However,
these models seem to over-predict the solar oxygen abundance derived
from helioseismology \citep{delahaye06}. Because of these
uncertainties, we have chosen to present our
results without any corrections.

Because of limited wavelength coverage and unfortunate gaps from
the pre-upgrade HIRES CCD, we are only able to measure the $\lambda$6707
line of lithium for three stars.  The lowest metallicity star with
a Li abundance is the dwarf CS 22884-108, at [Fe/H]$=-3.13$, for which we find log$\epsilon$(Li)$=2.31$. The
other two measurements are for the stars CS 22872-102 and CS22878-027,
with log$\epsilon$(Li) values of 1.97 and 2.39, respectively. The
estimated error for all three measurements is 0.17, and the
average value for these three metal-poor dwarfs is 2.21. This agrees
very well with the value reported by \citet{bonifacio07}, whose study
of 17 metal-poor dwarfs find an average log$\epsilon$(Li)$=2.10$. In
Figure \ref{cno}, it is clear that our values fall on the Spite
plateau, while the giants from \citet{spite05} show a significant
amount of depletion.

\subsection{Odd-Z elements: Na, Al, and Sc \label{oddz}}

In 10 of our stars observed with the original HIRES CCD, our setup
allowed us to measure Na from the Na D resonance lines at 5890 and
5895 \AA{}, and we were also able to measure Al in all but eight of our stars
using one or both of its resonance lines at 3944 and 3961 \AA. As
noted by \citet{cayrel04}, using these features introduces possible
NLTE effects. We report our abundances without NLTE corrections. The
suggested NLTE corrections when using these lines are $-0.5$ for Na
\citep{nanlte} and +0.65 for Al \citep{alnlte}.

\begin{figure}
\begin{center}
\scalebox{.4}[.4]{
\plotone{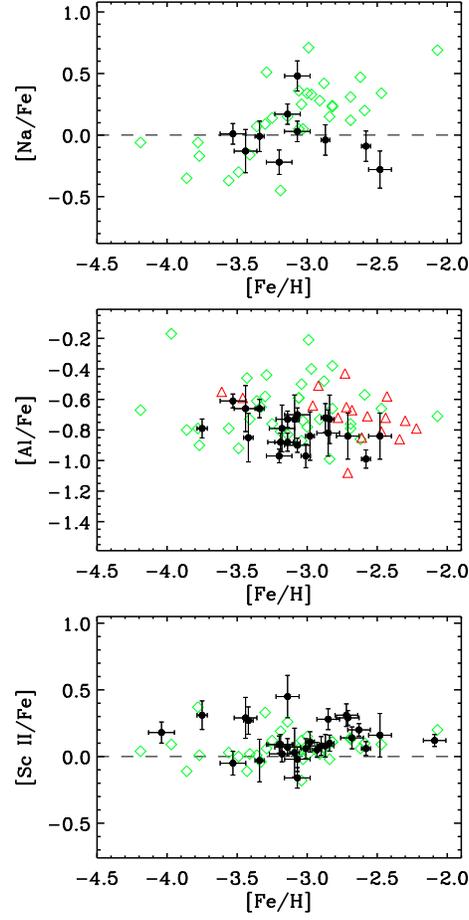}}
\end{center}
\figcaption{[(Na,Al,Sc)/Fe] vs. [Fe/H].  The diamonds
  (colored green in the electronic edition)
  are from \citet{cayrel04} and the triangles (colored red in the
  electronic edition) are from
  \citet{cohen04}. Although we only have two stars below [Fe/H] of
  $-2.7$ with measured Na, we do not find the trend found by
  \citet{cayrel04}. The [Al/Fe] values from \citet{cohen04} are
  plotted without the NLTE correction of 0.6 dex assumed in that
  study. See the electronic edition of the Journal for a color
  version of this figure.\label{naalsc}}
\end{figure}

As can be seen from Figure \ref{naalsc}, we do not reproduce the same
trend in [Na/Fe] versus [Fe/H] as \citet{cayrel04}. Although we have
fewer measurements of [Na/Fe] in the higher metallicity range, our
data suggest a flat trend in [Na/Fe]. Taken with the
\citet{cayrel04} points, there appears to be a large scatter in
[Na/Fe] down to [Fe/H]$\sim-3.25$, and then little scatter for lower
metallicities.

Figure \ref{naalsc} also shows that our [Al/Fe] abundances agree
remarkably well with the \citet{cayrel04} and \citet{cohen04}
distribution of values. In the metallicity range covered by this
study, we find a very low 0.12 dex dispersion in [Al/Fe].

\subsection{Alpha elements} 
\begin{figure}
\begin{center}
\scalebox{.4}[.4]{
\plotone{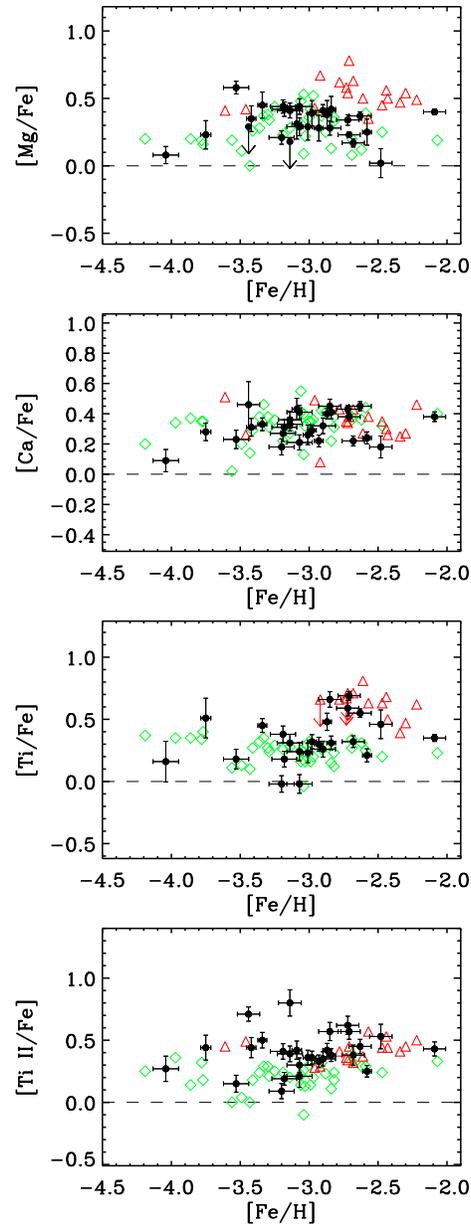}}
\end{center}
\figcaption{[$\alpha$/Fe] vs. [Fe/H].  The symbols are as
  in figure \ref{naalsc}. See the electronic edition of the Journal for a color
  version of this figure. \label{alpha}}
\end{figure}

The scatter in our $\alpha$-elements, as shown in Fig. \ref{alpha}, is
very small over the entire range of our metallicities. The rms scatter
of these elements is very similar to the value found by previous
studies working in our metallicity regime. There are no truly
anomalous outliers to the expected $\alpha$-enhancement trend
(e.g. \citealt{aoki07b,cohen07}). This suggests that all of our stars
formed from gas produced with a very similar star formation history.

For [Mg/Fe] and [Ca/Fe] the observed scatters are 0.13 and 0.1
with average values of 0.32 and 0.31, respectively. We also find
\ion{Ti}{2} to be on average 0.09 higher than \ion{Ti}{1}. However both the
neutral and singly ionized species of Ti show relatively large scatter, about
0.17 dex for both. We caution using this as a sign of
true abundance scatter as we find a correlation of Ti with \teff{}. We
discuss this further in $\S$ \ref{trends}.

\begin{figure}
\begin{center}
\scalebox{.5}[.5]{
\plotone{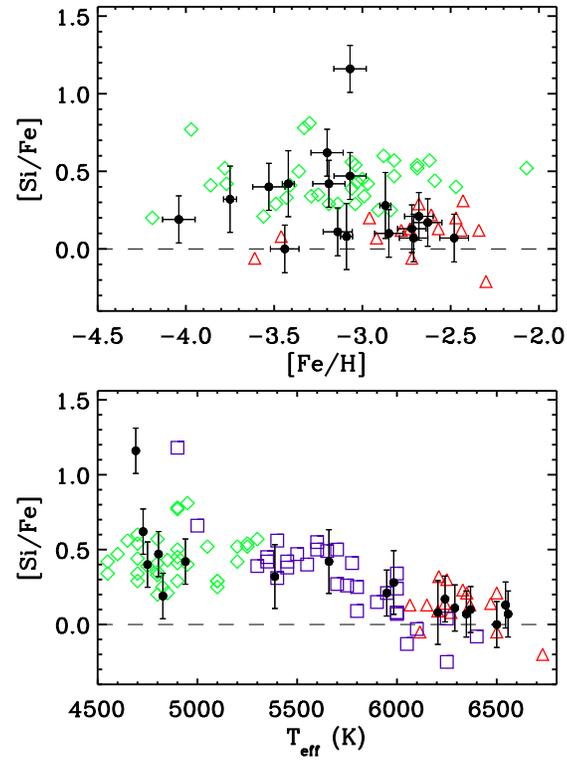}}
\end{center}
\figcaption{[Si/Fe] vs. [Fe/H] and \teff{}. The symbols are as
  in figure \ref{naalsc}. In the bottom panel, we also add the data
  from \citet{preston06} as the squares (colored purple in the
  electronic edition). See the electronic edition of the Journal for a color
  version of this figure.
\label{si}}
\end{figure}

Another element that exhibits a large scatter is Si. As we show in
Figure \ref{si}, we do not find any trends with metallicity and find a
rms of 0.24 dex in [Si/Fe]. It appears that part of this scatter comes from
a correlation of Si abundance with \teff{} which we also show in
Figure \ref{si}. This is also discussed further in $\S$ \ref{trends}.

\subsection{The Fe group (23$\leq$Z$\leq$28)}

\begin{figure}
\begin{center}
\scalebox{.4}[.4]{
\plotone{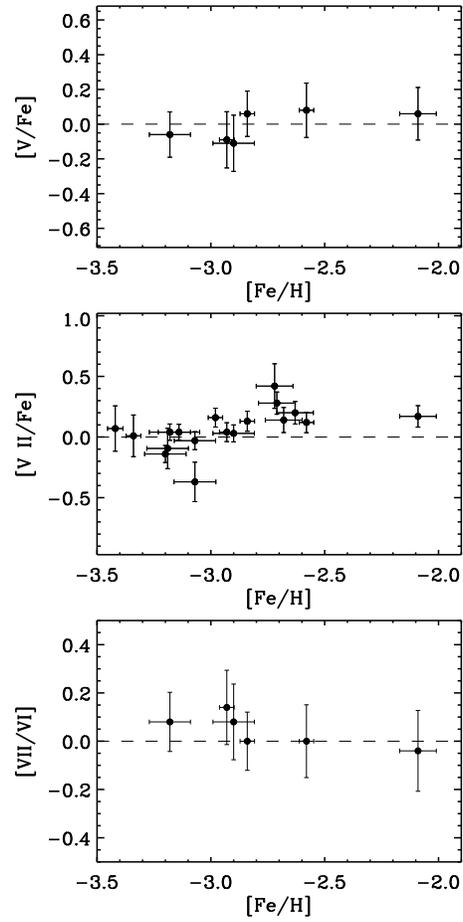}}
\end{center}
\figcaption{Vanadium abundance vs. [Fe/H]. We plot both the neutral and
  singly ionized species, along with [V II/V I]. We find no offset
  between the V II and V I abundances. 
\label{vanadium}}
\end{figure}

Figure \ref{vanadium} shows the $V$ abundances
for our sample. Although we do not measure \ion{V}{1} for many of our stars, we
do not find an offset between \ion{V}{2} and \ion{V}{1} as found by
\citet{johnson2002}. Overall both the neutral and ionized species
give no trend with [Fe/H], reflecting the similar origins of $V$ and Fe
from explosive silicon and oxygen burning \citep{ww95}

\begin{figure}
\begin{center}
\scalebox{.4}[.4]{
\plotone{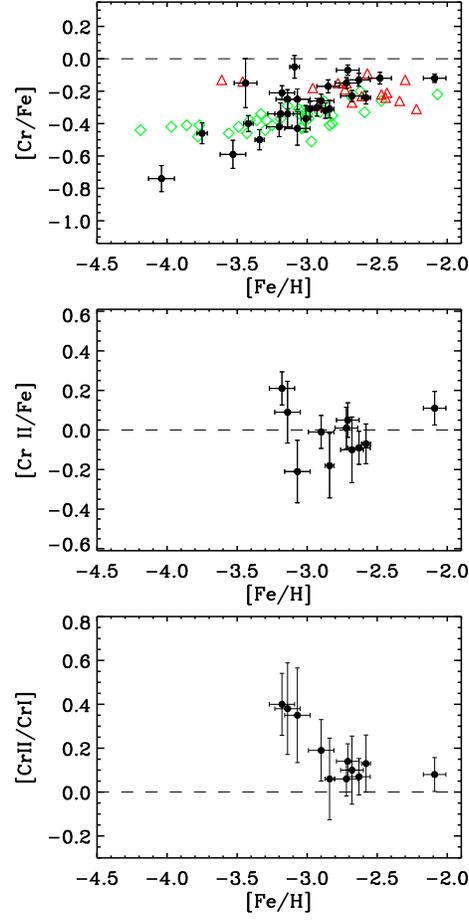}}
\end{center}
\figcaption{Cr abundance vs. [Fe/H]. We plot both the neutral and
  singly ionized species, along with [Cr II/Cr I]. The symbols are as in
  figure \ref{naalsc}. Although we do not measure Cr II for many of our
  stars, our results suggest an offset from zero for [Cr II/Cr I], as
  well as an increasing trend in this value with decreasing
  metallicity. See the electronic edition of the Journal for a color
  version of this figure.
\label{cr}}
\end{figure}

In Figure \ref{cr}, we plot both \ion{Cr}{1} and \ion{Cr}{2} as a function of
[Fe/H]. We reproduce the declining trend of [\ion{Cr}{1}/Fe] with [Fe/H] as
found in \citet{cayrel04} and references therein, albeit with a
slightly steeper slope. Although we are
only able to measure [\ion{Cr}{2}/Fe] for our more metal-rich objects
([Fe/H]$>-3.3$), we find a clear offset between the species, with
an average [\ion{Cr}{2}/\ion{Cr}{1}] of 0.22. Furthermore in the metallicity range
$-3.3<$[Fe/H]$<-2$ that we measure \ion{Cr}{2}, we find no evidence for
declining [\ion{Cr}{2}/Fe] with declining [Fe/H]. We explore this more in
$\S$ \ref{trends}.

\begin{figure}
\begin{center}
\scalebox{.4}[.4]{
\plotone{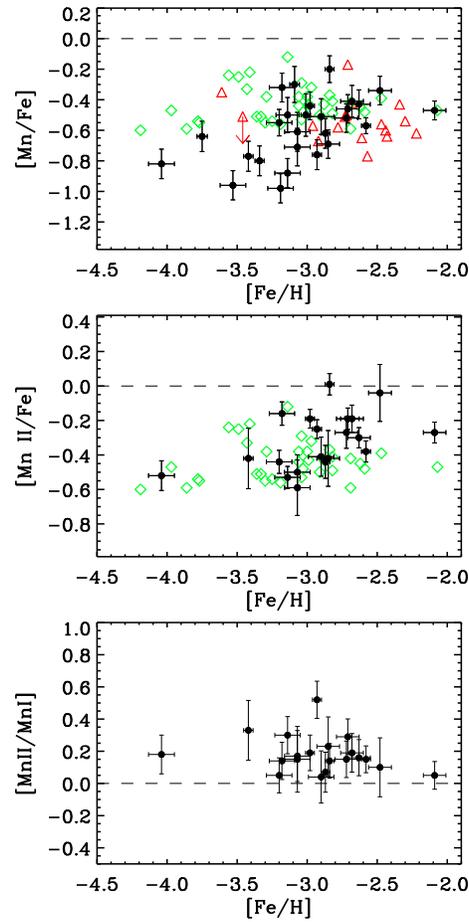}}
\end{center}
\figcaption{Manganese abundance vs. [Fe/H]. We plot both the neutral and
  singly ionized species, along with [Mn II/Mn I]. The symbols are as in
  figure \ref{naalsc}. In the [Mn II/Fe]
  plot, we also overplot the [Mn/Fe] values from \citet{cayrel04} for
  the reasons described in the text. See the electronic edition of the Journal for a color
  version of this figure.
\label{manganese}}
\end{figure}

Figure \ref{manganese} similarly summarizes our Mn abundances. The
differences between our study and that of \citet{cayrel04}
arise from a correction that they chose to adopt for the abundances
determined from the \ion{Mn}{1} resonance triplet at 4030 \AA{}. They note
that in their objects these lines give a consistently lower value
($\sim0.4$ dex) for Mn than the non-resonance lines. They therefore do
not include them in the final Mn abundance. In the stars where only
the triplet lines were detected, their abundance was adopted with a
correction of +0.4 dex. We have chosen to include the \ion{Mn}{1}
triplet in our abundance determinations without
correction, since it may have some unknown dependence on the
atmospheric parameters of a star (the \citealt{cayrel04} sample's
concentration on giants avoids this problem). Whatever the cause
of this discrepancy, it makes the \ion{Mn}{1} abundances suspect. For
this reason, we concentrate on the \ion{Mn}{2} abundances. In Figure
\ref{manganese} we have also plotted the \ion{Mn}{1} abundances from
\citet{cayrel04} on our \ion{Mn}{2} abundances. As can be seen from the
figure, our agreement is quite good, although we are finding slightly
higher Mn abundances at [Fe/H]$>-3.0$. 

\begin{figure}
\begin{center}
\scalebox{.5}[.5]{
\plotone{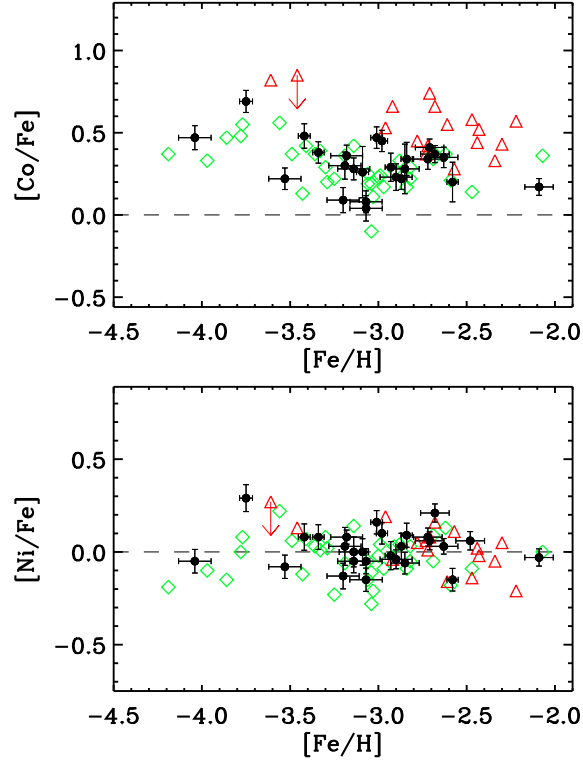}}
\end{center}
\figcaption{[Co,Ni/Fe] vs. [Fe/H]. The symbols are as in
  figure \ref{naalsc}. We find a similar trend
  of increasing [Co/Fe] with decreasing [Fe/H] as in
  \citet{cayrel04}. The [Ni/Fe] shows no trend with [Fe/H]. See the electronic edition of the Journal for a color
  version of this figure.
\label{coni}}
\end{figure}

We generally confirm the increase in [Co/Fe] with decreasing [Fe/H] as
found by \citet{mcwilliam95} and \citet{cayrel04}. In Figure
\ref{coni} we see that the trend matches the points from
\citet{cayrel04} in general scatter and slope. It is unclear why the
points from \citet{cohen04} are offset from ours. The line parameters
for Co are similar among all three studies. Our [Ni/Fe] values are
also shown in this figure, and they stay flat across all
metallicities, with a very low rms of 0.10 dex. 

\subsection{Cu and Zn}

\begin{figure}
\begin{center}
\scalebox{.5}[.5]{
\plotone{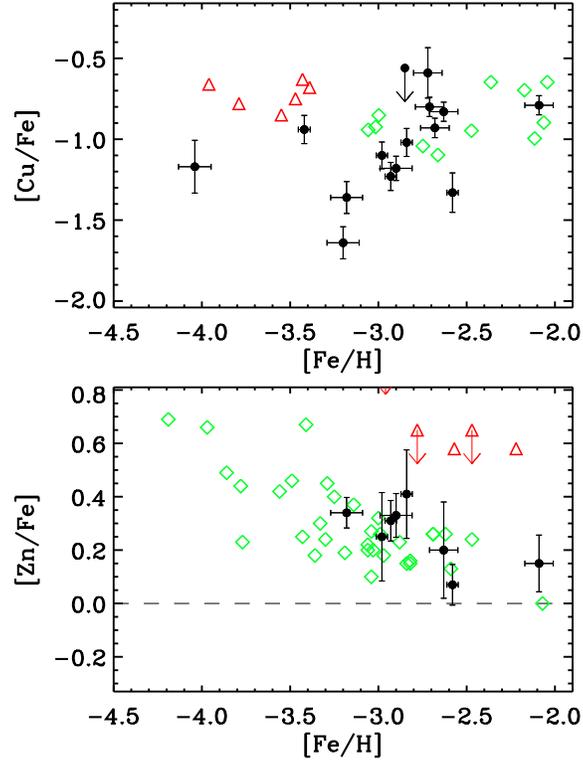}}
\end{center}
\figcaption{[(Cu,Zn)/Fe] vs. [Fe/H]. The triangles (colored
  red in the electronic edition) in the
  [Cu/Fe] plot are from \citet{cohen07-2}, and the diamonds (colored
  green in the electronic edition) are
  from \citet{bihain}. In the [Zn/Fe] plot, the symbols are as in
  figure \ref{naalsc}. See the electronic edition of the Journal for a color
  version of this figure.\label{cuzn}}
\end{figure}

In Figure \ref{cuzn}, we plot [Cu/Fe] versus [Fe/H] and [Zn/Fe]
versus [Fe/H]. It has been known since \citet{sneden88} that Cu is
deficient in metal-poor stars. We are able to extend the [Cu/Fe]
measurements down to [Fe/H]$=-4.0$, comparable to the metallicities of
\citet{cohen07-2}, and down from the previous low of [Fe/H]$\sim-3.0$
(e.g. \citealt{mishenina02,simmerer03,bihain}).

The abundance determinations from
\citet{cohen07-2} and \citet{bihain} are also shown in the [Cu/Fe]
plot. While we find a similar distribution of [Cu/Fe] values as
\citet{bihain}, \citet{cohen07-2} measure higher [Cu/Fe]
abundances. Part of this abundance spread is likely artificial. The
version of MOOG used by \citet{cohen07-2}, incorrectly treats continuum
scattering as absorption. This is properly accounted for in the
Turbospectrum code (e.g. \citealt{cayrel04}). We also derived [Cu/Fe] using MOOG and found that
those values are on average 0.22 dex higher, with as much as a 0.6 dex
difference. Our values are in good agreement with the chemical
evolution models of \citet{romano} which assume an initial primary
origin of Cu in SNeII, and a secondary contribution from the weak
$s$-process.

We are only able to measure [Zn/Fe] in a handful of our stars due to
gaps in the wavelength coverage of some of our spectra. The values
that we are able to measure agree nicely with the results of
\citet{cayrel04}. In particular, our data are consistent with the rise
of [Zn/Fe] with decreasing [Fe/H]. Taken together, this rise in [Zn/Fe]
could be indicative of an $\alpha$-rich freeze out process
contributing at a higher level at low metallicities \citep{cayrel04}.

\subsection{Neutron capture elements \label{ncaptsec}}

\begin{figure}
\begin{center}
\scalebox{.4}[.4]{
\plotone{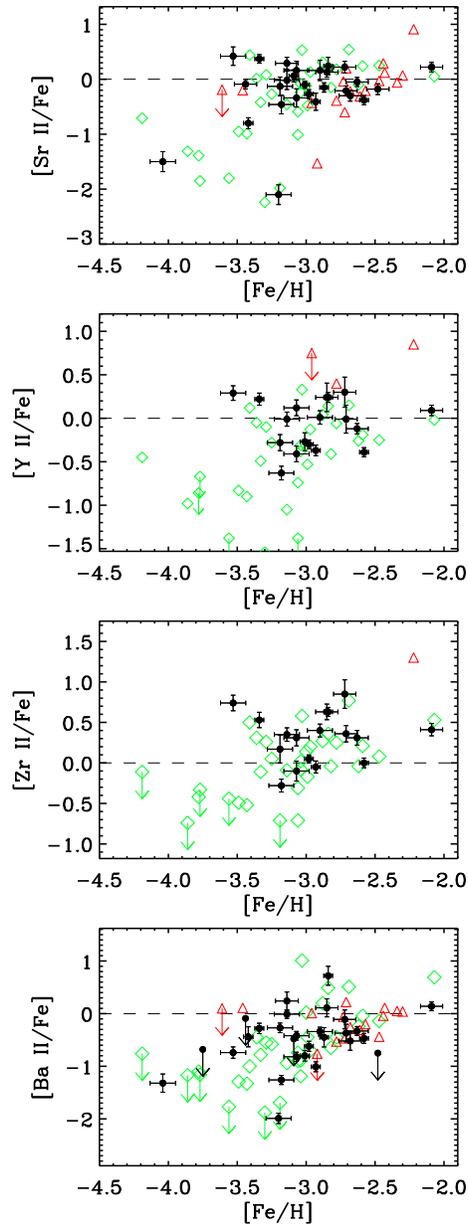}}
\end{center}
\figcaption{[(Sr II, Y  II, Zr II, Ba II)/Fe] vs. [Fe/H]. We clearly
  see a large scatter in all of these abundances. The triangles
  (colored red in the electronic edition) are
  from \citet{cohen07-2} and the diamonds (colored green in the
  electronic edition) are from \citet{francois}. See the electronic edition of the Journal for a color
  version of this figure.
\label{ncaptplotorig}}
\end{figure}

We are able to measure four neutron-capture elements in most of our
stars: the light peak elements Sr, Y, and Zr ($Z=$38, 39, and 40) and
the heavy neutron capture element Ba (Z=56). We find a large scatter
in these elements relative to Fe, as can be seen in Figure
\ref{ncaptplotorig}.

\begin{figure}
\begin{center}
\scalebox{.4}[.4]{
\plotone{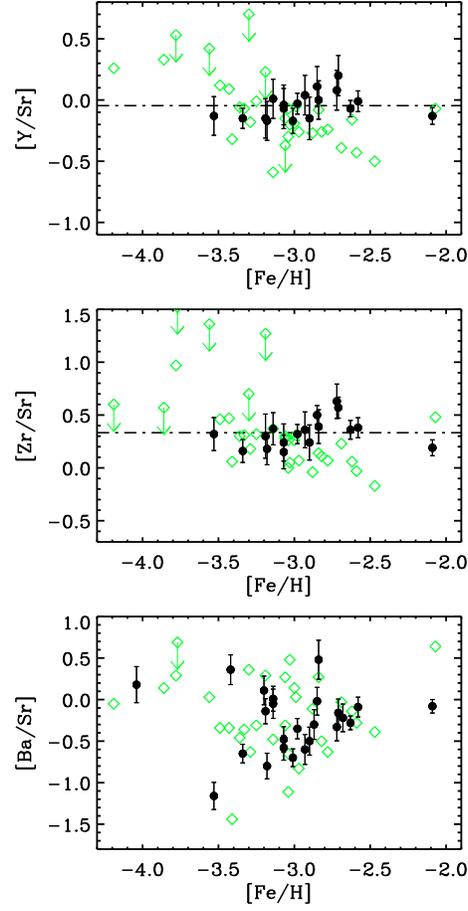}}
\end{center}
\figcaption{The light neutron-capture elements, Sr, Zr,
  and Y, show remarkable correlation, while [Sr/Ba] shows a scatter of
  almost 2 dex. The diamonds (colored green in the electronic edition) are from \citet{francois}. See the electronic edition of the Journal for a color
  version of this figure.
\label{eltosr}}
\end{figure}

As found by other studies (e.g., \citealt{mcwilliam95,jb2002,honda04,aoki05,francois}), the light
elements also appear to be highly correlated with each other,
suggesting that these elements share a common origin. In Figure
\ref{eltosr} we show that [Y/Sr] and [Zr/Sr] have consistent values 
for all stars with their respective averages of $-0.05$ and 0.33.

The Ba resonance line abundances agree well with the non-resonance
lines when measurable, and while we only have three stars with La measurements,
these values are roughly consistent with the Ba measurements. The
light neutron-capture elements, however, show a remarkable scatter
relative to Ba. Figure \ref{eltosr} also shows [Ba/Sr] versus
[Fe/H]. The difference between the extreme values is almost 2 dex,
replicating the finding of previous work
\citep{mcwilliam98,jb2002,honda04}.

\begin{figure}
\begin{center}
\scalebox{.75}[.75]{
\plotone{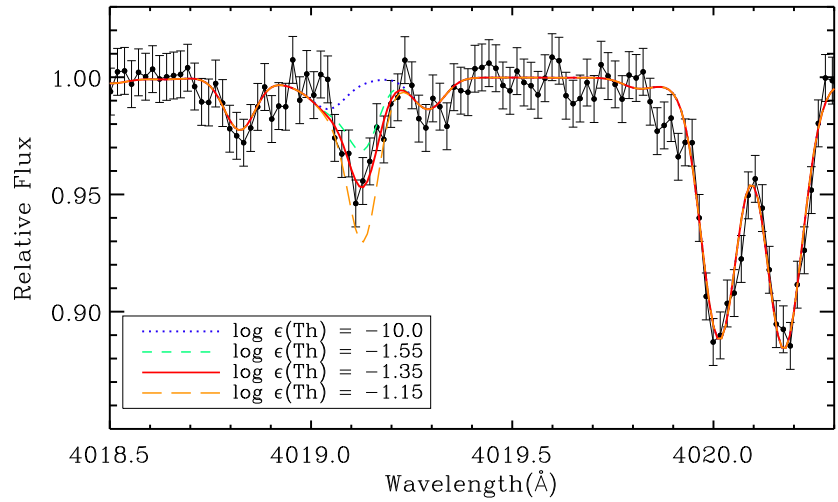}}
\end{center}
\figcaption{Spectral synthesis of the Th line at 4019 \AA{}
  in CS 31078-018. See the electronic edition of the Journal for a color
  version of this figure.
\label{th4019}}
\end{figure}

We also report on the discovery of a new $r$-process-rich star, CS 31078-018.
Table \ref{ncapt} gives the summary of its Z$\geq38$ abundances,
including thorium. In Figure \ref{th4019} we show the spectral
synthesis of the $\lambda$4019 line of \ion{Th}{2}, using the log $gf$ value
measured by \citet{nilsson02}. For most of the lines used to measure
these neutron-capture elements hfs is negligible because of their low
EWs ($<$30 m\AA{}). Except for Ba and Eu the only line that is
noticeably affected in CS 31078-018 is the resonance line used to
measure Yb. To account for this we follow the prescription of \citet{sneden03}
and use the hfs parameters from \citet{ybhfs}.

\section{Discussion}

\subsection{Mixing}

\begin{figure}
\begin{center}
\scalebox{.4}[.4]{
\plotone{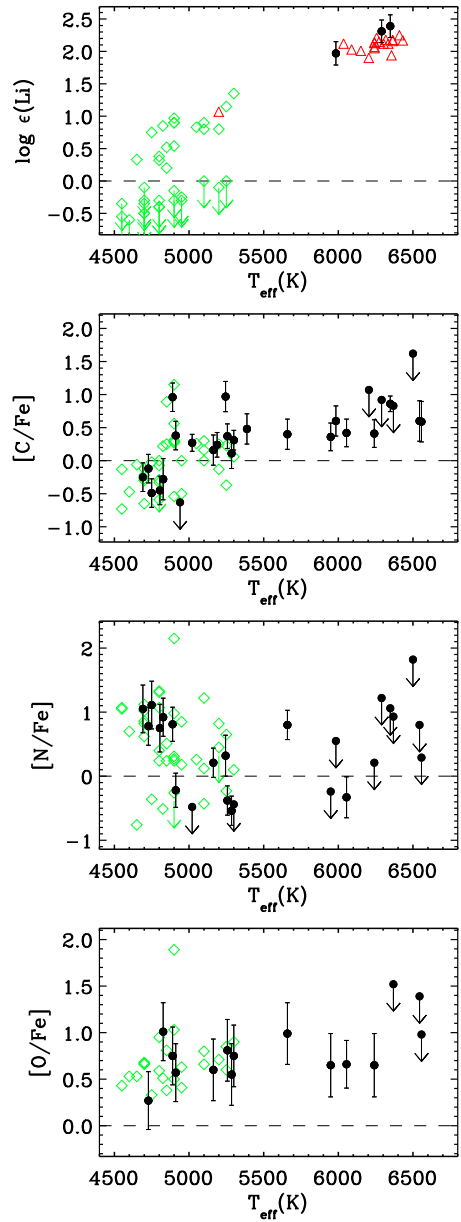}}
\end{center}
\figcaption{Values of log$\epsilon(\mbox{Li})$ and [C, N, O/Fe]
  vs. \teff{}. The symbols are as in figure \ref{cno}. See the electronic edition of the Journal for a color
  version of this figure. \label{light-teff}}
\end{figure}

\begin{figure}
\begin{center}
\scalebox{.4}[.4]{
\plotone{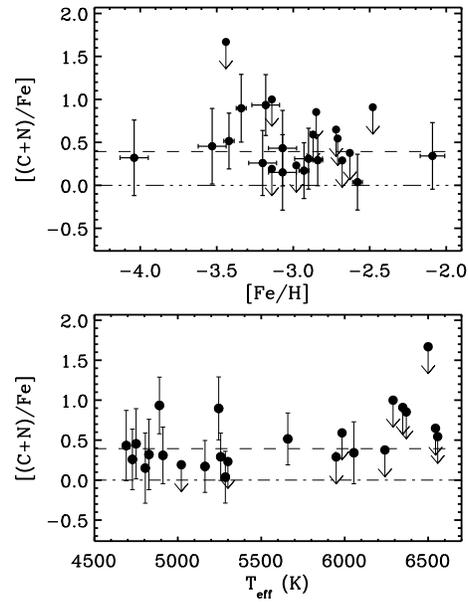}}
\end{center}
\figcaption{[(C+N)/Fe] vs. [Fe/H] and \teff{}. The large
  number of upper limits obscures any potential trend with [Fe/H];
  however, it seems clear that [(C+N)/Fe] is not correlated with
  \teff{}. The rms scatter for the measured [C+N/Fe] is 0.27 dex, and
  the average value is 0.39 dex.
\label{c+n}}
\end{figure}

For some elements, we expect differences between giants and dwarfs
because of internal mixing in giants. Models of mixing and
observations of the effect on the abundances in the atmospheres of
giants have a long history (e.g., \citealt{kraft94}). The most relevant observations for
very metal-poor giants are the \citet{gratton00} and \citet{spite05}
results. The mixing effects can be clearly seen in Figure
\ref{light-teff}, where we have plotted the light elements (C, N, O,
and Li) as a function of \teff{}, where we note that [C/Fe] drops 
off below $\sim$5000K. We also plot [(C+N)/Fe], which
should remain largely unchanged with evolution, as a
function of metallicity and \teff{} in Figure \ref{c+n}. There is no
correlation of [(C+N)/Fe] with [Fe/H] or with \teff{}, at least for the
region for which we have more than upper limits, \teff{} below 5700K.

\begin{figure}
\begin{center}
\scalebox{.5}[.5]{
\plotone{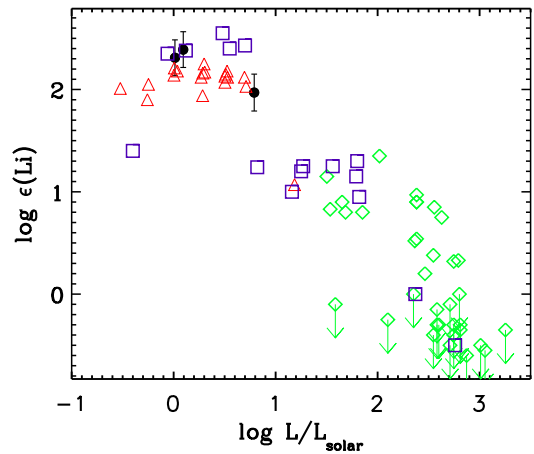}}
\end{center}
\figcaption{Values of log$\epsilon$(Li) vs.luminosity. The symbols are
  as in figure \ref{cno}, with the addition of the squares (colored purple in
  the electronic edition) from \citet{gratton00}. See the electronic edition of the Journal for a color
  version of this figure.\label{lithium}}
\end{figure}

In Figure \ref{lithium} we plot log$\epsilon$(Li) as a function of
luminosity for our sample as well as those of \citet{gratton00} (for
[Fe/H]$<-1.3$), \citet{spite05}, and \citet{bonifacio07}. We see the
clear signature of the deepening convective envelope. The average
log$\epsilon$(Li) for the three dwarfs with Li measurements is 2.21,
consistent with the Spite plateau value of 2.10 as found by
\citet{bonifacio07}.

\subsection{Si, Ti and Cr, unexpected trends with \teff{} \label{trends}}

Other than for the elements discussed above, mixing processes and
therefore the stage of stellar evolution that a star is in should not
affect its abundances. However for the Si, Ti, and Cr abundances we
do find a correlation of abundance with \teff{}. 

We plot our values of [Si/Fe] versus \teff{} in Figure \ref{si}.  It
is clear from this figure that there is a trend of decreasing Si with
increasing \teff{}. The \citet{preston06} study of field horizontal branch
stars and red horizontal branch stars in M15 also suggests this
result, and shows that it is not correlated with \logg{} as well. This
puzzling trend is unexpected from an evolutionary standpoint; the
lack of correlation with \logg{} shows that it is not tied to mixing
along the red giant branch. \citet{preston06} carefully checked for
contamination from possible CH lines and found very little effect, so
that is unlikely the culprit. 

\begin{figure}
\begin{center}
\scalebox{.4}[.4]{
\plotone{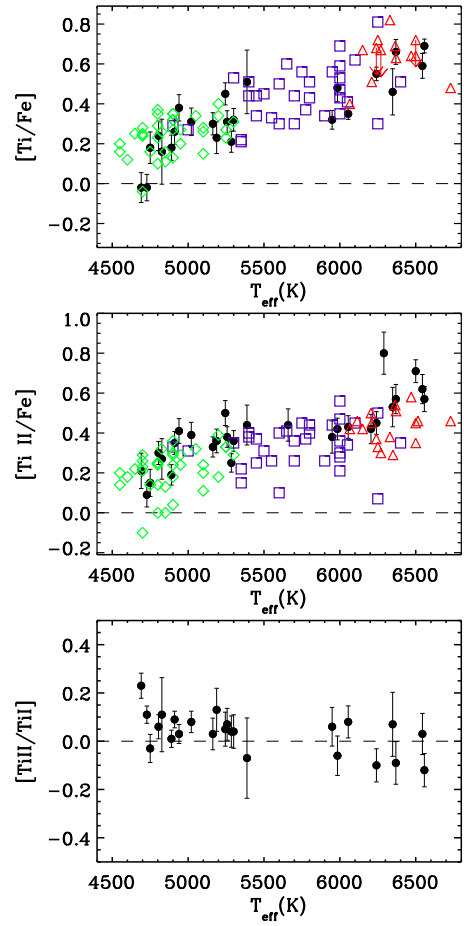}}
\end{center}
\figcaption{[Ti/Fe] and [Ti II/Fe] versus \teff{}. We also
  plot [Ti II/T I], which shows that, although there is an offset
  between these values, the trend exists for both the neutral and
  singly ionized states. We also plot data taken from
  \citet{preston06} as the squares (colored purple in the electronic
  edition). See the electronic edition of the Journal for a color
  version of this figure.
\label{titrend}}
\end{figure}

We also find a trend of Ti abundance versus \teff{} in the opposite
sense of Si. As we show in
Figure \ref{titrend}, this trend shows a decrease in Ti with
decreasing \teff{}. In Figure \ref{titrend} the points
from \citet{cayrel04} and \citet{cohen04} seem to confirm this
trend. We have also included the data from
\citet{preston06} to fill in the sparse region in \teff{} between 5400
and 6000 K. With the inclusion of these horizontal branch stars from \citet{preston06}, we see that
like Si, this is predominately a \teff{}, not \logg{}
correlation. It also should be noted that this trend applies for both
\ion{Ti}{1} and \ion{Ti}{2}, with respective slopes of 0.12 dex in [\ion{Ti}{1}/Fe] per 500K and 0.10 dex
in [\ion{Ti}{2}/Fe] per 500K.

\begin{figure}
\begin{center}
\scalebox{.4}[.4]{
\plotone{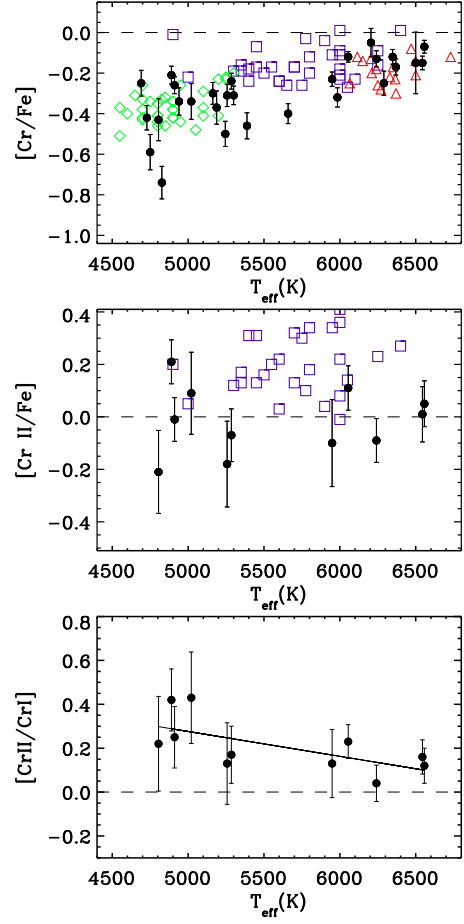}}
\end{center}
\figcaption{[Cr/Fe] and [Cr II/Fe] vs. \teff{}. We also
  plot [Cr II/Cr I], which shows that although there is an offset
  between these values, the trend exists for both the neutral and
  singly ionized states. The symbols are as in figure \ref{titrend}.
\label{crtrend}}
\end{figure}

A similar trend to Ti is found in \ion{Cr}{1}. In Figure \ref{crtrend} we plot
\ion{Cr}{1} and \ion{Cr}{2} as a function of \teff{}. A trend of declining
[Cr/Fe] with \teff{} can be seen, while [\ion{Cr}{2}/Fe] appears, if not
flat, then far less pronounced. Again we include data from
\citet{preston06} in our plots, and we come to the same conclusion
that this is a \teff{} and not a \logg{} correlation. The slopes of
the best fit lines for our data are 0.09 dex in [Cr/Fe] per 500K and
only 0.03 dex in [\ion{Cr}{2}/Fe] per 500K. This discrepancy in both slope
and offset between \ion{Cr}{1} and \ion{Cr}{2} may point to NLTE
effects. This has already been suggested by
\citet{sobeck07}, and the case seems to be made even stronger by our
data. Because \ion{Cr}{2} appears relatively free of \teff{} trends, we
performed a linear fit of [\ion{Cr}{2}/\ion{Cr}{1}] with \teff{} and suggest
that at least as a first step to correct \ion{Cr}{1} abundances by this
fit. We find that at 6500K, there is an offset of 0.106 dex and a
slope of -0.0113 dex per 100 K. This leads to a correction of 0.3 dex
at 4800K.

A possible explanation for the observed trends with \teff{} is an
incorrect $T$-$\tau$ relationship in the adopted model atmospheres, in
particular in giants. In fact, the adoption of a $T$-$\tau$ relationship
shallower than the true one would result in a derived abundance
dependent on the depth of the line formation and hence on its
strength, with strong lines yielding larger abundance values than weak
lines. While this effect on the derived [Fe/H] abundances can be, at
least partially, compensated by adjusting the value of the
micro-turbulence, this does not apply to Si, Cr and Ti, as the
abundances for the first come prevalently from lines forming the outer
layers and that for the other two mostly from lines formed deep into
the atmosphere. The observed trends could thus be explained in terms
of increasing discrepancy between the model atmospheres and the "true"
atmospheres at the decreasing of the stellar effective temperature. It
is noteworthy that the derived slopes for the $\chi$-derived abundance
relationship are steeper among giants than among dwarfs (see Table
\ref{atm}), which is what is expected in the hypothesis described.

Whether these are ultimately due to true abundance trends, unaccounted
for blends, or a deficiency in our knowledge in the spectral analysis
(e.g. NLTE and other atmosphere effects) is yet to be
determined. However by including the data from \citet{preston06}, it
seems that this is not an effect of stellar evolution. Regardless,
these trends show that caution must be taken when looking at either
Ti, \ion{Si}{1} or \ion{Cr}{1} to explore galactic chemical evolution
or to constrain SN models.

\subsection{CS 30336-049, [Fe/H]=$-4.0$}

Of our sample, the star CS 30336-049 is the most metal poor at
[Fe/H]=$-4.0$.  There are only three stars more metal-poor than this,
and only about seven with comparable metallicities (see
\citealt{frebel07b}). It is comparable in metallicity and atmospheric
parameters to both CD-38:245 and CS 22949-037 \citep{cayrel04}. In its
abundance ratios, CS 30336-049 looks far more like CD-38:245 than CS
22949-037.  Unlike for CS 22949-037, the derived [Mg/Fe] is actually
slightly under-abundant at 0.08 compared to other metal-poor stars.
While the high C,N,O and Mg of CS 22949-037 may be suggestive of a
low-energy SN explosion \citep{tsuj2003}, CS 30336-049 is
``normal'' in all of these abundances except for a high [N/Fe],
leading to a low [C/N] of $-1.2$. At the temperature range of this
star, $\sim$4800K, deep mixing may begin to occur and change some of
the C to N. Although some of the N then may come from this internal
processing in the star, it is possible that much of the low [C/N]
value comes from the initial abundances of the cloud that formed the
stars.  The low carbon abundance of CS 30336-049, coupled with its
very low [Ba/Fe] and [Sr/Fe] suggests that the star is not showing
mass transferred material from an AGB companion, which suggests a
primary source for its enhanced nitrogen abundance.

Just comparing CS 30336-049 with CS 22949-049, it is clear that
finding more stars at [Fe/H]$=-4$ is very important. Including
HE 1424-0241 as discussed in the introduction, the picture gets even
more complicated. If these are all products of one to a small number
of SN events, then a census of these objects may provide the best way
to constrain the nature of Population III stars. With this in mind, we have
attempted to fit our abundances to the recent SN II model yields of
\citet{heger08} in $\S$ \ref{fits}.

The low-metallicity of CS 30336-049 also makes it a good candidate for
constraining the gas cooling mechanism responsible for the Population III to
Population II transition. As noted by \citet{frebel07c} and references
therein, there are two main competing ideas for this cooling mechanism
following the initial metal enrichment from the first stars: atomic
fine-structure line cooling and dust-induced
fragmentation. Accordingly, \citet{frebel07c} define a value they term
as the ``transition discriminant,'' $D_{\mbox{trans}}$, that is
dependent on the overall C and O abundance of a star. They find that
the cooling from \ion{C}{2} and \ion{O}{1} fine-structure lines
can allow for low-mass star formation only at values of
$D_{\mbox{trans}}>-3.5 \pm 0.2$. Intriguingly, CS 30336-049 has a
$D_{\mbox{trans}}$ value of -3.57. Given the inherent uncertainty of
deriving O from the UV OH lines and the possibility that some C has
been converted to N, this is not a highly certain value. However, if this
value is correct, then this may be an indication that this star was
formed due primarily to fine-structure line cooling from C and O
produced from a Population III star. We note that this does not exclude the
dust-induced fragmentation model, as even lower abundances of C and O
can induce low-mass star formation in this scenario, but that now
including CS 30336-049 all metal-poor stars with C and O measurements
are consistent with the fine-structure cooling scenario.

\subsection{Neutron-capture-poor stars}

The existence of stars that are highly neutron-capture element deficient has
been known for some time
(e.g. \citealt{mcwilliam95,ryan96,mcwilliam98}). However these confirmed
neutron-capture poor stars are relatively rare. In our own sample, BS 16084-160
and CS 30336-049 are exceptional in their low [Sr/H] and [Ba/H]
values. Both of these values for BS 16084-160 agree with those
reported by \citet{aoki05}. \citet{fulbright04} analyzed the Draco
dSph red giant D119 and found upper limits to [Sr/H] and [Ba/H] that
match our values. The recent studies of
\citet{cohen07-2}, \citet{francois}, \citet{honda04}, and \citet{aoki05} have also added
to a handful of stars in this regime. In Table \ref{ncaptlow} we
summarize measurements from recent high-resolution studies for stars
with both [Ba/Fe]$<-1.0$ and [Ba/H]$<-4.0$.

\begin{figure}
\begin{center}
\scalebox{.5}[.5]{
\plotone{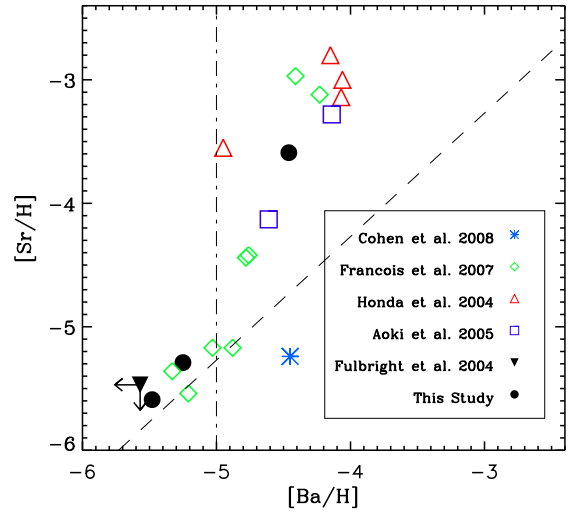}}
\end{center}
\figcaption{[Sr/H] vs. [Ba/H]. In these
  neutron-capture deficient objects, it seems that there are at least
  two sites that produce [Sr/H]. The main $r$-process line, as given in
  \citet{simmerer04} is plotted as the dashed line. See the electronic edition of the Journal for a color
  version of this figure.
\label{ncaptplot}}
\end{figure}

In these neutron-capture element-deficient stars, a striking pattern emerges
when comparing [Ba/H] to [Sr/H]. In Figure \ref{ncaptplot} we can see
that there seem to be two populations of stars. Below
[Ba/H]$\sim-5.0$, the Sr and Ba abundances appear to be well
coupled. Above this value, however, there is a clear distribution of
production of Sr relative to Ba, with the trend of Sr being mainly
over-produced relative to Ba. This suggests that the same process,
such as a very low-level main $r$-process, is producing both the Sr and
Ba in the hyper-neutron capture-poor stars (HNCP, [Ba/H]$<-5.0$).

To explain the production of Sr without much Ba, leading to high
[Sr/Ba] ratios, \citet{qian2001} and \citet{travaglio} propose a
separation of the process that creates the light (e.g. Sr) and heavy
(e.g. Ba) neutron-capture elements (termed the light elementary
production process, LEPP, by \citealt{travaglio}). It has been proposed that the
production site for these elements may come from charged-particle
reactions in the neutrino driven wind off of a newly born neutron star
(\citealt{qian2007} and references therein). Recently both
\citet{montes} and \citet{qian2007} explored how a two-component
process could account for much of the scatter found in the light to
heavy neutron-capture abundances. Figure \ref{ncaptplot} shows that at
the HNCP end this LEPP may somehow be shut off, leaving only the
signature of the main $r$-process. While there has been much
concentration on neutron-capture rich stars, the observational
constraints on the production site of the light neutron-capture
elements will benefit greatly from more of these HNCP stars being
discovered.

\subsection{CS 31078-018, A new $r$-process-rich star}

\begin{figure}
\begin{center}
\scalebox{.8}[.8]{
\plotone{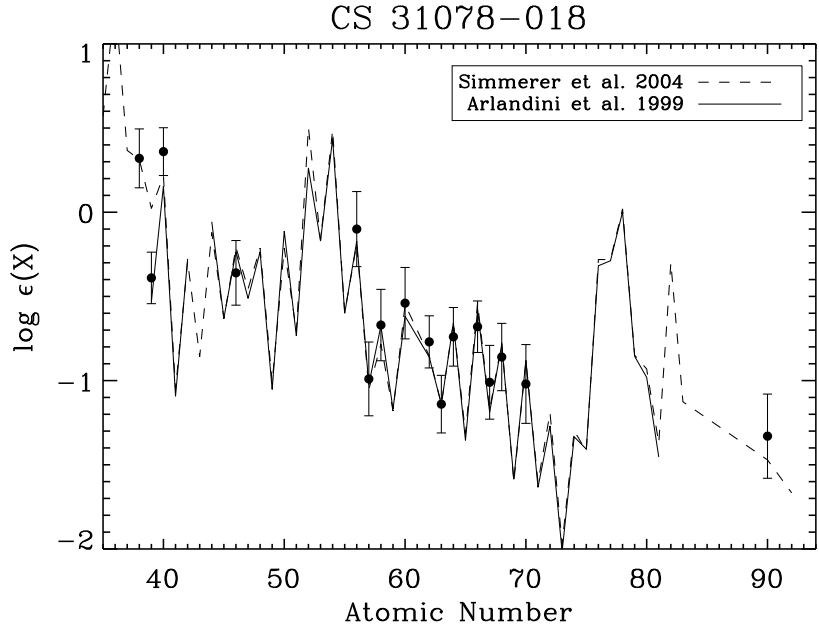}}
\end{center}
\figcaption{Measured neutron-capture elements in the
  star CS 31078-018. The solar system $r$-process lines come from
  \citet{arlandini} and \citet{simmerer04}, and have been scaled to
  match our Eu abundance.
\label{r-process}}
\end{figure}

We have discovered a new $r$-process-rich star, CS 31078-018. Adopting
the categories of \citet{beers05}, CS 31078-018 is an $r$-II star,
that is, [Eu/Fe]$>1$ and [Ba/Eu]$<0$. In Figure \ref{r-process} we plot
the neutron-capture abundances of this star over the solar system
$r$-process abundances taken from \citet{arlandini} and
\citet{simmerer04}.

Similar to previously discovered $r$-process-rich stars, the agreement
between the solar system $r$-process pattern and the abundances for CS
31078-018 is quite good for Z$\geq$56, further strengthening the case
for a universal ``main'' $r$-process for the stable neutron-capture
elements.

We also find that CS 31078-018 exhibits an ``actinide boost.'' The Th value
is far higher than what should be expected, given its radioactive
decay lifetime and the theoretical initial $r$-process production
ratio. Our measured value of log$\epsilon$(Th/Eu)$=-0.19$ gives a
negative age for the star if compared to the current estimate of
$-0.28$ for the initial $r$-process production ratio
\citep{kratz07}. This is the third star to exhibit this behavior,
after CS 31082-001 \citep{hill02} and CS 30306-132
\citep{honda04}.

In Fig. \ref{thorium} we show the available measurements of
log$\epsilon$(Th/Eu) from the literature for metal-poor stars. There
are some clear uncertainties as shown by some of the disagreements
between measurements for the same star, but it appears that there is a
real distribution of values of log$\epsilon$(Th/Eu), with CS
31078-018 near the top. The discrepancy between some of the results
from \citet{honda04} and other measurements is at least partially
explained by differences in their linelist with \citet{westin}, and
differences in adopted atmospheric parameters with
\citet{jb01}. \citet{qian2002} proposes a mechanism that allows for
a universal $r$-process site for the heavy $r$-nuclei but can vary the
actinide abundance via neutrino-induced fission, with neutrino
exposure being dependent on individual SNe II. A distribution of
values may lend credence to the idea of neutrino-induced fission
modifying the abundance of thorium. On the other hand, there seems to
be an almost bimodal distribution of values, which is even more
enhanced if we do not include the values from \citet{honda04}. This is
suggestive of two distinct scenarios with values typified by CS
22892-052 and CS 31078-018. Ultimately more [Th/Eu] measurements will
be needed to settle this question.

Regardless of the mechanism that over-produces the
actinides in some stars, it is clear from Fig. \ref{thorium} that it
does not affect all $r$-process-enhanced stars. Aside from this actinide
boost, there is no significant difference between these stars and
normal $r$-process-enhanced stars, and chronometers based on comparing
actinides to these stable elements must be approached with caution.

\subsection{Abundance Trends and Scatter in the Metal-poor Galaxy}

\begin{figure}
\begin{center}
\scalebox{.8}[.8]{
\plotone{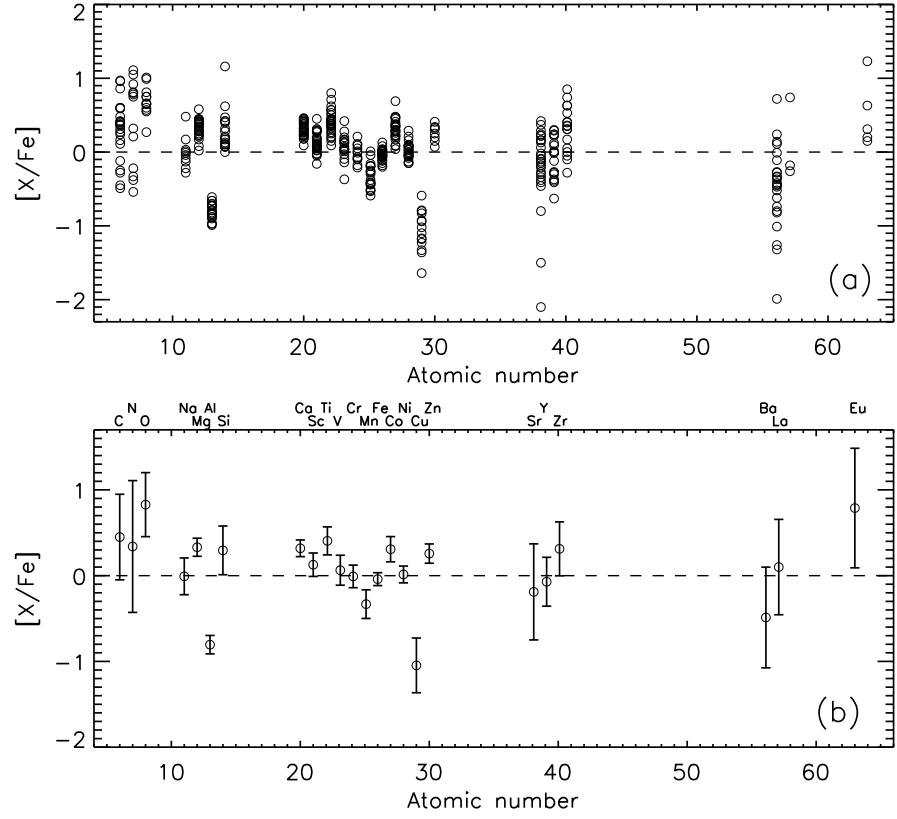}}
\end{center}
\figcaption{Abundance spread of our sample. When
  both the ionized and neutral species of an element are measured, we
  plot only the results for the ionized species. The points plotted
  for Fe are the \ion{Fe}{2} - \ion{Fe}{1} values of our stars. ($a$) Here we plot all
  of our measurements for the sample from C through Eu. ($b$) Instead of
  all of the measurements, we show the average measured abundance for
  each element. The error bars represent the rms of the abundances
  of each respective element.
\label{elemall}}
\end{figure}

Fig. \ref{elemall} summarizes the abundance measurements for our
entire sample. In the metallicity range that we cover, the only elements to
show a trend with [Fe/H] are Cr, Mn, Co, and Cu; our points are
consistent with \citet{cayrel04} for the trend they find in Zn. These
abundance ratio trends are reflected in their larger scatters in Figure \ref{elemall}. As
mentioned before, some of the scatter for \ion{Cr}{1} may be artificial since
we also see a trend with \teff{}, which also can explain why
\citet{cayrel04} find a much smaller scatter in \ion{Cr}{1} as they have
a much narrower \teff{} range. 

There is a true abundance spread in C and N, although as Figure
\ref{c+n} shows, some of this can be reconciled by considering
[(C+N)/Fe] (and therefore evolutionary state) instead. The large
scatter in the neutron-capture elements, particularly below [Fe/H]$=-3$
as shown in Figure \ref{ncaptplotorig} is also readily apparent in \ref{elemall}.

What stands out from Figure \ref{elemall}, however is not just the
scatter of the previously mentioned elements, but the remarkable
consistency in other elements. For example the values for [Ca/Fe],
[V/Fe], and [Ni/Fe], given reasonable errors, are effectively the same
for all of our stars. This may also be true of [Si/Fe], [Ti/Fe], and
[Cr/Fe], but this may be masked by their \teff{} dependence. We
are seeing either the evidence of a very well mixed ISM when these stars
formed, or evidence that the progenitor stars, possibly Population III stars,
were all very similar.

\subsection{Abundance Pattern fits to zero metallicity SN \label{fits}}

If we assume that these early stars were formed from the products of
the first stars, then we can try matching them to some of the most
up-to-date nucleosynthesis results from zero-metallicity SN
explosions. We use the recent models of \citet{heger08}, which are
updates to the models of \citet{ww95}. These models range in mass from
10 to 100 M$_{\odot}$, energies from 0.3 to 10.0 ($\times10^{51}$
ergs), and mixing from none to 0.25. \citet{heger08} use a
1D code and mixing cannot be followed due to its
multi-dimensional nature. Instead, in these models an artifical
prescription for this parameter is used from \citet{pinto88}, and the
amount of mixing is defined in terms of the mass fraction of the
helium core (for a detailed explanation see \citealt{heger08}). We then assume that
these stars are the product of one to just a few SN (e.g.,
\citealt{tumlinson06}) or assume an IMF of the first stars and match
the yield from that to our stars.

We first fit the average abundances of our sample to these
models. There are a few general assumptions that we are using to guide
us. For all of our fits we are assuming that the Sc and Zn
abundances derived from the models are lower limits. It is possible
that part of the abundances of both of these elements are synthesized in
proton-rich outflows from core-collapse SNe \citep{pruet}, and part of
the zinc may also be made in a neutrino-powered wind
\citep{hoffman96}. Neither of these processes is included in the
models of \citet{heger08}. We fit for C+N, instead of C and N
separately to account for potential internal cycling, and we ignore O
because of possible offsets due to using the OH feature. Finally we
have added in the NLTE corrections discussed in $\S$ \ref{oddz} to our
Na and Al abundances. 

We also find a curious result that the models
always over-produce copper relative to the observations. In general we
ignore copper in our fits, although we discuss it more below. We are
also concerned with the trends of Ti, Si, and \ion{Cr}{1} discussed in $\S$
\ref{trends}, as well as the offset between the neutral and ionized
species of Mn. For Ti and Si, we weight their values by only half in
the fits compared to the other elements. For Cr and Mn, we choose to use the
averages from their ionized species to reduce the impact of potential NLTE effects
on their respective neutral species.

In the IMF models the explosion energy of the SN for a star of mass
$M$ is parameterized as
\begin{equation}
E=E_0 \times \left(\frac{M}{20\,\mathrm{M}_{\odot}}\right)^{E^{\mathrm{exp}}}\,.
\label{energyeq}
\end{equation}
As a convenient SN energy unit we use bethe, $1\,\mathrm{B}=
10^{51}\,$ergs. Possible values of $E_0$ in the model database are 0.3, 0.6, 0.9, 1.2, 1.5, 1.8, 2.4, 3.0, 5.0, and
10.0 B, and $E^{\mathrm{exp}}$ can be $-1, -0.5, 0$, and 1.

\begin{figure} \begin{center}
\scalebox{.5}[.5]{ 
\plotone{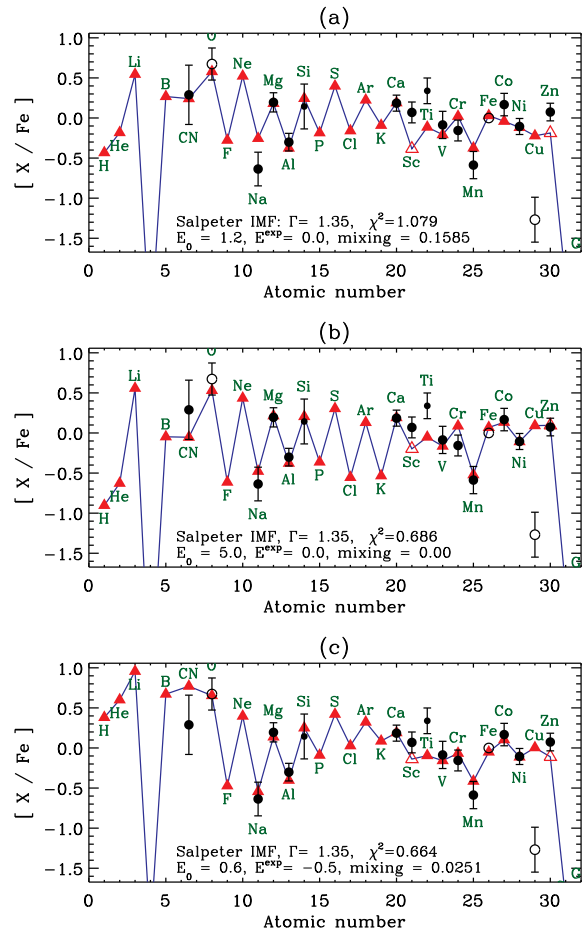}}
\end{center}
\figcaption{Average abundance pattern of our
  sample fitted to the \citet{heger08} models assuming a
  Salpeter IMF. The filled circles are the averaged abundances, with
  the error bars corresponding to the rms of the abundance ratios over
  our sample. The smaller filled circles of Si and Ti represent the
  smaller weights attributed to them in the fitting procedure. The
  open circles are when that particular abundance is not used in the
  fit, and the open triangles at Sc and Zn represent treating the
  model yields as lower limits.
\label{salpall}}
\end{figure}

In Figure \ref{salpall}(a) we show the best fit for a Salpeter IMF
($\Gamma$=1.35), for progenitor masses from 10 to 100 M$_{\odot}$ (the
entire mass range of the models). In the first case, we assume a
standard IMF, where we set $E^{\mathrm{exp}}=0$ and $E_0=1.2$. The
only free parameter is the mixing fraction. The fit is
actually quite good, returning a standard mixing parameter of 0.16,
compared to the canonical value of 0.1 used to explain the light curve
of SN1987A (\citealt{pinto88}).

We relax the constraints on $E^{\mathrm{exp}}$ and $E_0$ in Figure
\ref{salpall}(b). We let the energy run over the ranges listed above,
and only restrict $E^{\mathrm{exp}}\geq0.0$, and we also let the progenitor lower
and upper masses float. The match is noticeably better, with
$\chi^2=0.69$ for the best fit model. The $E_0=5.0$B, with the
explosion energy flat over all masses, $E^{\mathrm{exp}}=0.0$, and no mixing. Also,
even though we let the lower and upper bounds of the IMF vary, the
best fit still used the entire mass range of models.

Finally in \ref{salpall}(c), we let all the parameters range
freely. This now includes letting $\Gamma$ have the values
$-0.65$, 0.35, 1.35, 2.35, and 3.35, to look at multiple Salpeter-like
IMFs. We find a nearly identical fit in terms of $\chi^2$ with
\ref{salpall}(b). The Salpeter power law exponent is still given by
$\Gamma$=1.35, and the mixing parameter is a low 0.025. The
characteristic energy is now much lower with 0.6B, and
$E^{\mathrm{exp}}=-0.50$. The low explosion energy coupled with the negative
value of $E^{\mathrm{exp}}$ means that we are in effect reducing the importance
of high-mass objects. The increased fallback in
these objects when the explosion energy is lowered (see
\citealt{zhang}) causes very little of their metals to be ejected, in
particular the innermost Fe-group elements fall back first.

\begin{figure} \begin{center}
\scalebox{.5}[.5]{ 
\plotone{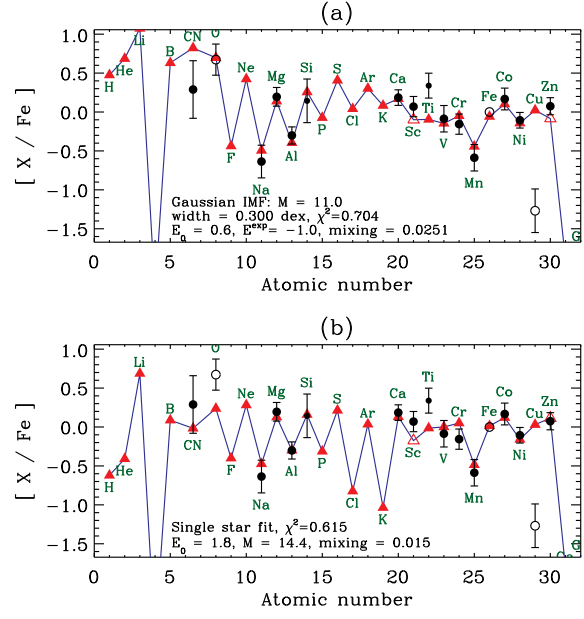}}
\end{center}
\figcaption{Average abundance pattern of our
  sample fitted to the \citet{heger08} models. The symbols are
  as in Fig. 26. ($a$) Best fit assuming a Gaussian
  IMF. ($b$) Best fit to a single SN.
\label{otherfits}}
\end{figure}

We also constructed Gaussian IMFs to try to match our average abundance
distribution. The Gaussians are centered at 11, 12, 13.5, 15, 17, 20,
25, 35, 40, 50, 75, and 100 M$_{\odot}$. The widths, in log(Mass),
range from 0.025 up to 0.5 dex. The energies of the explosions are
determined by Eqn. \ref{energyeq}, with the same ranges of $E_0$ and
$E^{\mathrm{exp}}$. The best-fit Gaussian IMF yield is shown in
Fig. \ref{otherfits}(a). The fit both looks qualitatively similar to
\ref{salpall}(c) and is also quantitatively very similar with a
best-fit $\chi^2=0.70$. This Gaussian IMF is centered at 11.0
M$_{\odot}$, with a width of 0.3 dex (truncated at 10.0
M$_{\odot}$). The energies are defined by $E^{\mathrm{exp}}=-1.0$ and
$E_0=0.6$, and the mixing is 0.0251. The similarity in the fits is not
surprising given that this truncated Gaussian IMF would look very
similar to the Salpeter IMF of \ref{salpall}(c). We also fit our
average abundance pattern to the entire library of single SN
yields. This may be instructive to define a ``typical'' Population III star,
even though clearly this is not the origin of the average
abundances. We show the result in Fig. \ref{otherfits}(b). The fit is
quite good, with a $\chi^2=0.615$. This best-fit star has a mass of
14.4 M$_{\odot}$, explosion energy of 1.8B, and a low mixing parameter
of 0.015.

\begin{figure}
\begin{center}
\scalebox{.5}[.5]{ 
\plotone{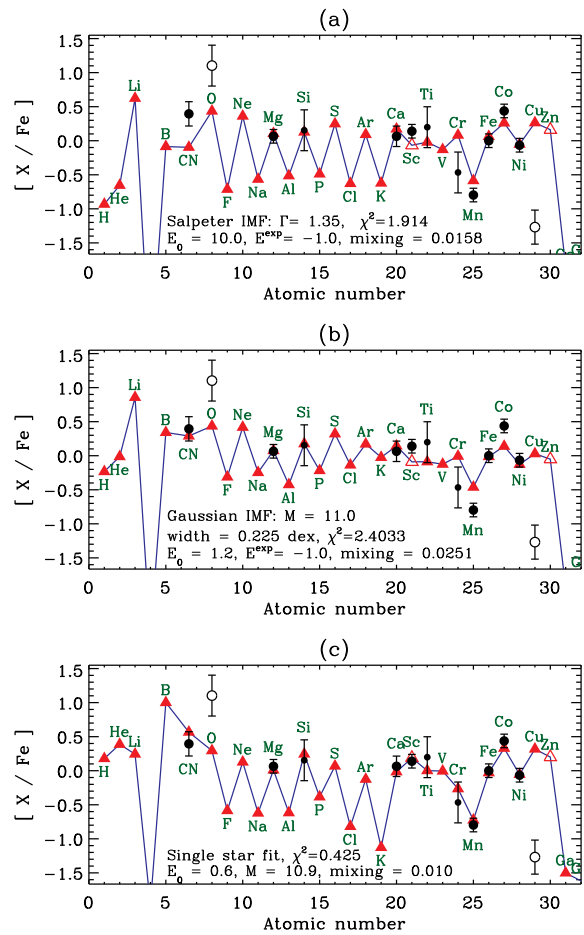}}
\end{center}
\figcaption{Abundance pattern of CS 30336-049 fitted
  to the \citet{heger08} models. Oxygen and copper are ignored,
  and the Cr has been increased by 0.3 dex. The type of fit and its
  parameters are listed on each plot.
\label{beststar}}
\end{figure}

We have also performed this best fit analysis on the most metal-poor
star of our sample, CS 30336-049. In Figure \ref{beststar} we show
cases similar to those described above, assuming a Salpeter IMF (a), a
Gaussian IMF (b), and a single star progenitor (c). We could not measure \ion{Cr}{2}
in this object, and instead we use the correction to \ion{Cr}{1} proposed in
$\S$ \ref{trends}, and adjust \ion{Cr}{1} by +0.3 dex for the fit. Because of
the inherent uncertainty of this abundance, we also weight Cr by a
factor of 1/2 as is done for Si and Ti. The other assumptions are as
detailed in the beginning of this section. The Salpeter IMF has a
difficult time fitting the abundance pattern of this star. The fit has
a $\chi^2=1.914$, a very high $E_0=10$B, $E^{\mathrm{exp}}=-1.0$, $\Gamma=1.35$,
mixing of 0.025, and ranges over all masses. The Gaussian IMF is best
fitted by $E_0=1.2$B, $E^{\mathrm{exp}}=-1.0$, a central mass of 11.0 M$_{\odot}$, a
width of 0.225 dex, and mixing of 0.025. The $\chi^2$ is also a somewhat
large 2.40. The most interesting case is the single star fit. The fit
is excellent at $\chi^2=0.425$, with M=10.9 M$_{\odot}$, E=0.6B, and a
very low mixing parameter of 0.01. This explosion in particular does an excellent
job of reproducing the Fe-peak element pattern.

\begin{figure}
\begin{center}
\scalebox{.5}[.5]{ 
\plotone{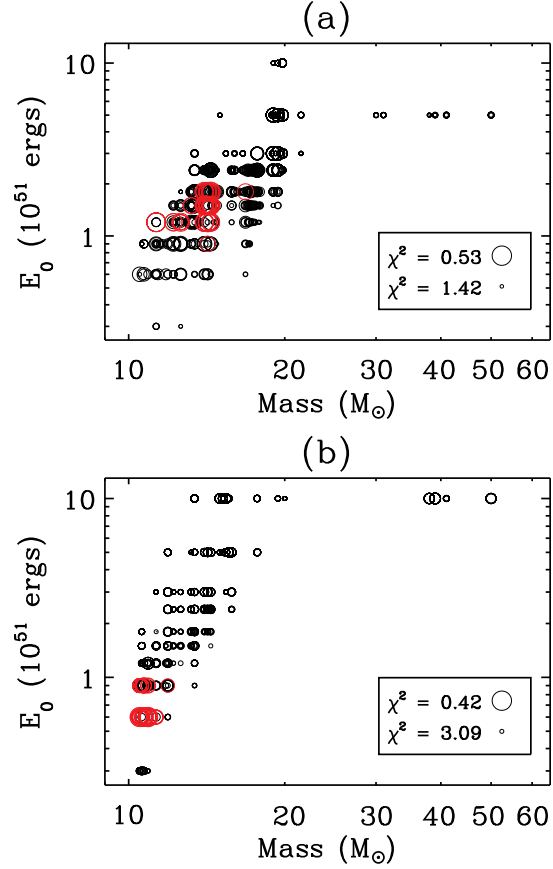}}
\end{center}
\figcaption{The 1000 best-fit single explosion models
  to ($a$) our average abundance pattern and ($b$) CS 30336-049. The $\chi^2$
  values are represented linearly by the size of the circles, with the
  minimum and maximum $\chi^2$ of the fits shown in the legends of each
  respective plot. Mixing values are not differentiated in these
  plots. The best 50 fits for each case are plotted in red.
\label{best1000}}
\end{figure}

As a final test, we also examined how well different parameters fit
our abundances, as opposed to looking only at the single best-fit
model. All of the best fits presented above favor progenitor stars
with fairly low characteristic masses, $\sim$ 10-15
M$_{\odot}$. The characteristic explosion energies, however, are not as
well constrained, with $E_0$ ranging from 0.6 to 5.0B. The reality is
that with so many models that can be compared to (16,800 total), it is
relatively easy to find very good fits that are also somewhat
degenerate in $\chi^2$. With this in mind we show in Figure
\ref{best1000} the best 1000 single SN fits in terms of $E_O$ and mass both
for the average abundance ratio of our sample and for CS
30336-049. The grid-like nature in these plots comes from the discrete
values of the models. Overall the results of the best fits from above
are borne out. Although in both cases there are a small number of models
that have very high explosion energy and mass, the vast majority of
fits for our average abundance pattern have masses in between 10 and
20 M$_{\odot}$ and $E_0$ less than 3.0B, while the results for the
abundance pattern of CS 30336-049 show an even narrower range of
masses of 10-15 M$_{\odot}$ and typical $E_0$ less than 1.0B. The
mixing parameter could not be constrained with this method because
mass and energy have far greater impact to the fits, and there is not
a favored mixing value when looking at the fits in aggregate. 

These results seem to fit well with the findings of
\citet{tumlinson06-2}, which suggests that the characterstic masses of
the first stars were between 8 and 42 M$_{\odot}$. This is argued from
empirical constraints based on the non-detection of Population III
stars in the Galactic halo, the Galactic halo MDF, and reionization,
not from detailed chemical abundances as done here. As
\citet{tumlinson06-2} notes, these numbers are also close to the
results of theoretical models of primordial star formation that
incorporate formation feedback effects (e.g. \citealt{brommlarson} and
references therein).

\subsubsection{Copper, A cautionary note on the choice of explosion mechanism}

It is unclear why copper is so over-produced in these
models. Part of the solution may come from the choice of these models
to locate the piston used to parameterize the explosion at an abrupt
entropy jump where the entropy per baryon ($S/N_Ak)$ is equal to
4, approximately at the base of the convective shell of the
pre-SN object. Previously in \citet{ww95}, it was located where the electron mole
number, $Y_{\mathrm{e}}$, decreased suddenly, which marked the edge of the iron core.

\begin{figure} \begin{center}
\scalebox{.5}[.5]{ 
\plotone{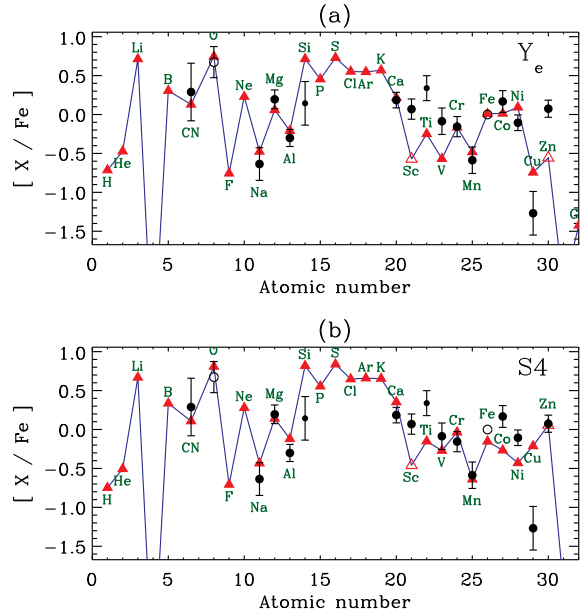}}
\end{center}
\figcaption{Plot of the effect of
  positioning the piston at the Y$_e$ boundary (edge of the iron core) and the S4 ($S/N_Ak=4.0$)
  boundary (base of the convective shell). The black points with error bars again represent the
  average abundance pattern of our sample. ($a$) Best single star
  fit assuming a piston location at the edge of the iron core. ($b$) Model with the same parameters, but with the piston location at the
  base of the convective shell.
\label{copperfit}
}
\end{figure}

In \citet{heger08}, there are also a small subset of models
with explosion energies of 1.2 and 10 B calculated with the piston
located at the $Y_{\mathrm{e}}$ boundary. To test the effect of the different
piston locations, we show in Fig. \ref{copperfit}(a) the best-fit
single-star model to our average abundances using the $Y_{\mathrm{e}}$ models
(including Cu). This model has a mass of 25.5 M$_{\odot}$, $E_0=1.2$B,
and mixing of 0.1. In \ref{copperfit}(b) we show their model with the
exact same parameters, but with the piston located at the $S/N_Ak$
boundary. It is clear that the location of the piston can greatly
affect the [Cu/Fe]. In the $Y_{\mathrm{e}}$ model [Cu/Fe] is about $-$0.70,
compared to the approximately solar value found in the same
$S/N_Ak$ model. This is not to say which explosion mechanism is
correct, as any specification is a parameterized approach, but it
does indicate that copper is a less than ideal element to use to
constrain current SN models.

\section{Summary}
We have presented the abundances from C to Eu of 28 metal-poor stars
covering a wide range of \teff{}. In the process we have found
abundance trends with \teff{} of \ion{Si}{1}, \ion{Ti}{1} and \ion{Ti}{2},
and \ion{Cr}{1} that may be pointing to the deficiencies of our
standard 1-D, LTE spectral analysis, or less likely, an unknown
physical process for these elements. In either case the unexplained trends in
\ion{Si}{1}, Ti, and \ion{Cr}{1} with \teff{} show that we must be
careful when using them to constrain models of galactic chemical
evolution and models of SN yields.

Our sample includes the discovery of a new [Fe/H]$=-4.0$ star, CS 30336-049. In
CS 30336-049, we find abundance ratios that track the trends from more
metal-rich objects ([Fe/H]$\sim-3.5$), except for a mildly low
[Mg/Fe]. These results for CS 30336-049 show that some of the stars around
[Fe/H]$=-4.0$ have a similar origin to these more metal-rich
objects. However, the 10 well-studied stars of similar or lower metallicity show a
diversity of abundances far greater than found in the more metal-rich stars.

\begin{figure}
\begin{center}
\scalebox{.5}[.5]{ 
\plotone{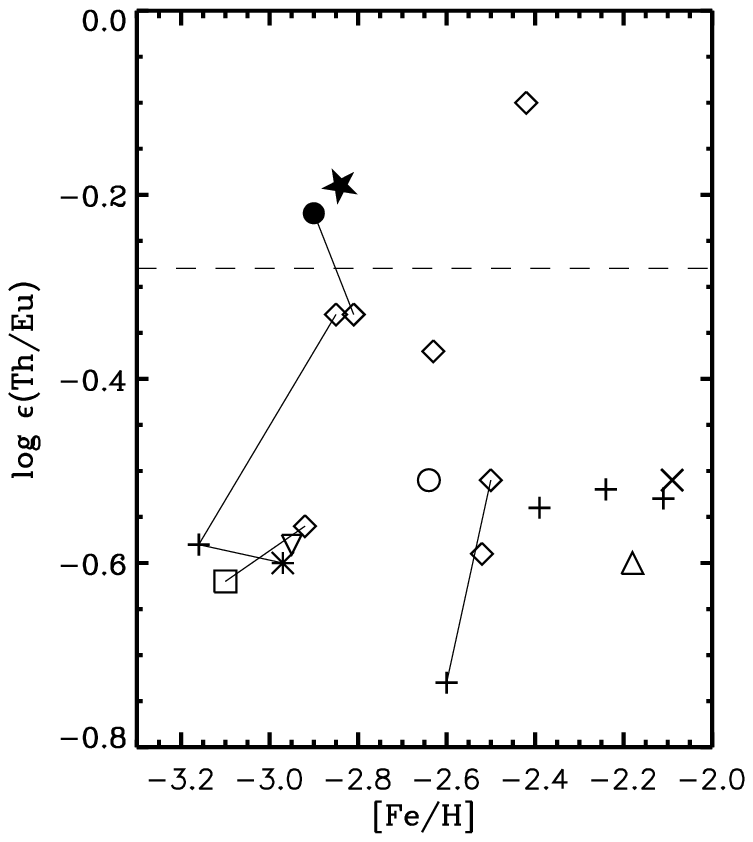}}
\end{center}
\figcaption{Measured log $\epsilon$(Th/Eu) abundances
  of metal-poor stars. The symbols represent the following: the filled star is CS 31078-018
  from this study, the diamonds are stars from \citet{honda04},
  the plus signs are stars from \citet{jb01}, the solid circle is
  CS 31082-001 from \citet{hill02}, the upward-pointing triangle is HD 221170
  from \citet{ivans06}, the downward-pointing triangle is HE 1523-0901
  from \citet{frebel07}, the square is CS 22892-052 from
  \citet{sneden03}, the cross is BD+17 3248 from \citet{cowan02},
  the open circle is CS 29497-004 from \citet{heres1},
  and the asterisk is HD 115444 from \citet{westin}. The solid
  lines connect points that are repeated measurements of the same
  object. \citet{honda04} suggest that the discrepancy in the
  measurements for HD 115444 arise from a combination of differing atmospheric
  parameters and linelists. There is a clear distribution of values,
  although a majority of the stars have log
  $\epsilon$(Th/Eu)$\sim-0.6$. The production ratio from \citet{kratz07} is plotted as the
  dashed line.
\label{thorium}}
\end{figure}

We have also discovered a new $r$-process enhanced star, CS
31078-018. Like other $r$-process-enhanced stars, it has a heavy
neutron-capture (Z$>$56) abundance pattern that matches the scaled
solar system $r$-process pattern. Interestingly, it has a much higher
[Th/Eu] than most other $r$-process-rich stars, one that matches the
value found in CS 31082-001. From figure \ref{thorium}, it is clear
that there is a diversity of [Th/Eu] values in $r$-process-rich
stars. Whether it is a bimodal distribution or continuous is not yet
clear, and it will take more Th measurements to be sorted
out. Regardless, it shows that chronometers based solely on this ratio
need to be used with caution.

We also explored the origin of the lighter neutron-capture elements
(Z$<56$) by examining stars that are highly deficient in these
elements. By using stars in this sample and the literature, we have
found that HNCP stars ([Ba/H]$\leq$5) only exhibit the
signature of the main $r$-process in their [Ba/Sr] abundance. This is in
contrast to stars with slightly higher [Ba/H], which show a wide
diversity of [Ba/Sr]. This result suggests that if there is a secondary
process that produces these lighter elements (i.e. Sr), then it does
not operate in the most metal-poor regime. In determining the secondary
physical process this may prove to be an important constraint if this 
continues to hold true as more HNCP stars are found.

Overall, we find very little scatter in our relative abundances for
elements in the Fe-group and lighter, and we have largely confirmed many
of the trends of abundance with metallicity for stars with
[Fe/H]$<-2.0$ that were detailed in \citet{cayrel04}. The low rms
suggests either a well mixed ISM or a common origin for our
stars. With this in mind, we have compared the average abundance
pattern of our sample with the zero-metallicity SN II nucleosynthesis
models of \citet{heger08}. These fits
seem to indicate that metal-free SN II progenitors with masses $\sim10
\mbox{-} 20$ M$_{\odot}$ can match our abundances very well. This
comparison was also done with the most metal-poor star of our sample,
CS 30336-049, where we find that a slightly narrower range of
progenitor masses $\sim10 \mbox{-} 15$ M$_{\odot}$ give the best matches to its
abundance pattern.

\acknowledgements

D.K.L. would like to acknowledge Chris Sneden, Ruth Peterson, Thomas
Masseron, Bob Kraft, and Yong-Zhong Qian for useful discussions, insights,
and advice.

D.K.L., M.B., and J.A.J. performed this work with the support of the National
Science Foundation (AST-0098617 and AST-0607770).

S.L. performed this work with the support of INAF cofin 2006 and the
DFG cluster of excellence ``Origin and Structure of the Uniferse.'' S.L. would
also like to thank R. Gratton for helpful discussion.

A.H. performed this work under the auspices of the US Department of
Energy at the University of California Los Alamos National Laboratory
under contract W-7405-ENG-36 and the DOE
Program for Scientific Discovery through Advanced Computing (SciDAC;
DE-FC02-01ER41176). S.W. received support from the National Science
Foundation (AST-02611) and
from the DOE SciDAC Program (FC02-06ER41438).

{\it Facilities:} \facility{Keck:I (HIRES)}

\bibliography{all.bib}

\begin{thebibliography}{107}
\expandafter\ifx\csname natexlab\endcsname\relax\def\natexlab#1{#1}\fi

\bibitem[{{Aldenius} {et~al.}(2007){Aldenius}, {Tanner}, {Johansson},
  {Lundberg}, \& {Ryan}}]{mgtriplet}
{Aldenius}, M., {Tanner}, J.~D., {Johansson}, S., {Lundberg}, H., \& {Ryan},
  S.~G. 2007, \aap, 461, 767

\bibitem[{{Alonso} {et~al.}(1996){Alonso}, {Arribas}, \&
  {Martinez-Roger}}]{alonso96}
{Alonso}, A., {Arribas}, S., \& {Martinez-Roger}, C. 1996, \aap, 313, 873

\bibitem[{{Alonso} {et~al.}(1999){Alonso}, {Arribas}, \&
  {Mart{\'{\i}}nez-Roger}}]{alonso99}
{Alonso}, A., {Arribas}, S., \& {Mart{\'{\i}}nez-Roger}, C. 1999, \aaps, 140,
  261

\bibitem[{{Alvarez} \& {Plez}(1998)}]{turbo}
{Alvarez}, R., \& {Plez}, B. 1998, \aap, 330, 1109

\bibitem[{{Anders} \& {Grevesse}(1989)}]{anders89}
{Anders}, E., \& {Grevesse}, N. 1989, \gca, 53, 197

\bibitem[{{Aoki} {et~al.}(2005){Aoki}, {Honda}, {Beers}, {Kajino}, {Ando},
  {Norris}, {Ryan}, {Izumiura}, {Sadakane}, \& {Takada-Hidai}}]{aoki05}
{Aoki}, W., {Honda}, S., {Beers}, T.~C., {Kajino}, T., {Ando}, H., {Norris},
  J.~E., {Ryan}, S.~G., {Izumiura}, H., {Sadakane}, K., \& {Takada-Hidai}, M.
  2005, \apj, 632, 611

\bibitem[{{Aoki} {et~al.}(2007){Aoki}, {Honda}, {Beers}, {Takada-Hidai},
  {Iwamoto}, {Tominaga}, {Umeda}, {Nomoto}, {Norris}, \& {Ryan}}]{aoki07b}
{Aoki}, W., {Honda}, S., {Beers}, T.~C., {Takada-Hidai}, M., {Iwamoto}, N.,
  {Tominaga}, N., {Umeda}, H., {Nomoto}, K., {Norris}, J.~E., \& {Ryan}, S.~G.
  2007, \apj, 660, 747

\bibitem[{{Arce} \& {Goodman}(1999)}]{arce99}
{Arce}, H.~G., \& {Goodman}, A.~A. 1999, \apjl, 512, L135

\bibitem[{{Arlandini} {et~al.}(1999){Arlandini}, {K{\"a}ppeler}, {Wisshak},
  {Gallino}, {Lugaro}, {Busso}, \& {Straniero}}]{arlandini}
{Arlandini}, C., {K{\"a}ppeler}, F., {Wisshak}, K., {Gallino}, R., {Lugaro},
  M., {Busso}, M., \& {Straniero}, O. 1999, \apj, 525, 886

\bibitem[{{Asplund} \& {Garc{\'{\i}}a P{\'e}rez}(2001)}]{asplund01}
{Asplund}, M., \& {Garc{\'{\i}}a P{\'e}rez}, A.~E. 2001, \aap, 372, 601

\bibitem[{{Barklem} {et~al.}(2005){Barklem}, {Christlieb}, {Beers}, {Hill},
  {Bessell}, {Holmberg}, {Marsteller}, {Rossi}, {Zickgraf}, \&
  {Reimers}}]{heres2}
{Barklem}, P.~S., {Christlieb}, N., {Beers}, T.~C., {Hill}, V., {Bessell},
  M.~S., {Holmberg}, J., {Marsteller}, B., {Rossi}, S., {Zickgraf}, F.-J., \&
  {Reimers}, D. 2005, \aap, 439, 129

\bibitem[{{Baumueller} {et~al.}(1998){Baumueller}, {Butler}, \&
  {Gehren}}]{nanlte}
{Baumueller}, D., {Butler}, K., \& {Gehren}, T. 1998, \aap, 338, 637

\bibitem[{{Baumueller} \& {Gehren}(1997)}]{alnlte}
{Baumueller}, D., \& {Gehren}, T. 1997, \aap, 325, 1088

\bibitem[{{Beers} {et~al.}(2000){Beers}, {Chiba}, {Yoshii}, {Platais},
  {Hanson}, {Fuchs}, \& {Rossi}}]{beers00}
{Beers}, T.~C., {Chiba}, M., {Yoshii}, Y., {Platais}, I., {Hanson}, R.~B.,
  {Fuchs}, B., \& {Rossi}, S. 2000, \aj, 119, 2866

\bibitem[{{Beers} \& {Christlieb}(2005)}]{beers05}
{Beers}, T.~C., \& {Christlieb}, N. 2005, \araa, 43, 531

\bibitem[{{Beers} {et~al.}(1992){Beers}, {Preston}, \& {Shectman}}]{bps}
{Beers}, T.~C., {Preston}, G.~W., \& {Shectman}, S.~A. 1992, \aj, 103, 1987

\bibitem[{{Bihain} {et~al.}(2004){Bihain}, {Israelian}, {Rebolo}, {Bonifacio},
  \& {Molaro}}]{bihain}
{Bihain}, G., {Israelian}, G., {Rebolo}, R., {Bonifacio}, P., \& {Molaro}, P.
  2004, \aap, 423, 777

\bibitem[{{Blackwell-Whitehead} {et~al.}(2005){Blackwell-Whitehead}, {Xu},
  {Pickering}, {Nave}, \& {Lundberg}}]{mnIupdate}
{Blackwell-Whitehead}, R.~J., {Xu}, H.~L., {Pickering}, J.~C., {Nave}, G., \&
  {Lundberg}, H. 2005, \mnras, 361, 1281

\bibitem[{{Bonifacio} {et~al.}(2007){Bonifacio}, {Molaro}, {Sivarani},
  {Cayrel}, {Spite}, {Spite}, {Plez}, {Andersen}, {Barbuy}, {Beers}, {Depagne},
  {Hill}, {Fran{\c c}ois}, {Nordstr{\"o}m}, \& {Primas}}]{bonifacio07}
{Bonifacio}, P., {Molaro}, P., {Sivarani}, T., {Cayrel}, R., {Spite}, M.,
  {Spite}, F., {Plez}, B., {Andersen}, J., {Barbuy}, B., {Beers}, T.~C.,
  {Depagne}, E., {Hill}, V., {Fran{\c c}ois}, P., {Nordstr{\"o}m}, B., \&
  {Primas}, F. 2007, \aap, 462, 851

\bibitem[{{Bonifacio} {et~al.}(2000){Bonifacio}, {Monai}, \&
  {Beers}}]{bonifacio00}
{Bonifacio}, P., {Monai}, S., \& {Beers}, T.~C. 2000, \aj, 120, 2065

\bibitem[{{Bromm} \& {Larson}(2004)}]{brommlarson}
{Bromm}, V., \& {Larson}, R.~B. 2004, \araa, 42, 79

\bibitem[{{Carretta} {et~al.}(2002){Carretta}, {Gratton}, {Cohen}, {Beers}, \&
  {Christlieb}}]{carretta02}
{Carretta}, E., {Gratton}, R., {Cohen}, J.~G., {Beers}, T.~C., \& {Christlieb},
  N. 2002, \aj, 124, 481

\bibitem[{{Castelli} \& {Kurucz}(2003)}]{castelli03}
{Castelli}, F., \& {Kurucz}, R.~L. 2003, in IAU Symposium, Vol. 210, Modelling
  of Stellar Atmospheres, ed. N.~{Piskunov}, W.~W. {Weiss}, \& D.~F. {Gray},
  20P--+

\bibitem[{{Cayrel} {et~al.}(2004){Cayrel}, {Depagne}, {Spite}, {Hill}, {Spite},
  {Fran{\c c}ois}, {Plez}, {Beers}, {Primas}, {Andersen}, {Barbuy},
  {Bonifacio}, {Molaro}, \& {Nordstr{\"o}m}}]{cayrel04}
{Cayrel}, R., {Depagne}, E., {Spite}, M., {Hill}, V., {Spite}, F., {Fran{\c
  c}ois}, P., {Plez}, B., {Beers}, T., {Primas}, F., {Andersen}, J., {Barbuy},
  B., {Bonifacio}, P., {Molaro}, P., \& {Nordstr{\"o}m}, B. 2004, \aap, 416,
  1117

\bibitem[{{Christlieb} {et~al.}(2004){Christlieb}, {Beers}, {Barklem},
  {Bessell}, {Hill}, {Holmberg}, {Korn}, {Marsteller}, {Mashonkina}, {Qian},
  {Rossi}, {Wasserburg}, {Zickgraf}, {Kratz}, {Nordstr{\"o}m}, {Pfeiffer},
  {Rhee}, \& {Ryan}}]{heres1}
{Christlieb}, N., {Beers}, T.~C., {Barklem}, P.~S., {Bessell}, M., {Hill}, V.,
  {Holmberg}, J., {Korn}, A.~J., {Marsteller}, B., {Mashonkina}, L., {Qian},
  Y.-Z., {Rossi}, S., {Wasserburg}, G.~J., {Zickgraf}, F.-J., {Kratz}, K.-L.,
  {Nordstr{\"o}m}, B., {Pfeiffer}, B., {Rhee}, J., \& {Ryan}, S.~G. 2004, \aap,
  428, 1027

\bibitem[{{Christlieb} {et~al.}(2002){Christlieb}, {Bessell}, {Beers},
  {Gustafsson}, {Korn}, {Barklem}, {Karlsson}, {Mizuno-Wiedner}, \&
  {Rossi}}]{christlieb02}
{Christlieb}, N., {Bessell}, M.~S., {Beers}, T.~C., {Gustafsson}, B., {Korn},
  A., {Barklem}, P.~S., {Karlsson}, T., {Mizuno-Wiedner}, M., \& {Rossi}, S.
  2002, \nat, 419, 904

\bibitem[{{Christlieb} {et~al.}(2000){Christlieb}, {Reimers}, {Wisotzki},
  {Reetz}, {Gehren}, \& {Beers}}]{christlieb00}
{Christlieb}, N., {Reimers}, D., {Wisotzki}, L., {Reetz}, J., {Gehren}, T., \&
  {Beers}, T.~C. 2000, in The First Stars: Proceedings of the MPA/ESO Workshop
  Held at Garching, Germany, 4-6 August 1999, ESO ASTROPHYSICS SYMPOSIA. ISBN
  3-540-67222-2. Edited by A. Weiss, T.G. Abel, and V. Hill. Springer-Verlag,
  2000, p. 49, ed. A.~{Weiss}, T.~G. {Abel}, \& V.~{Hill}, 49--+

\bibitem[{{Cohen} {et~al.}(2002){Cohen}, {Christlieb}, {Beers}, {Gratton}, \&
  {Carretta}}]{cohen02}
{Cohen}, J.~G., {Christlieb}, N., {Beers}, T.~C., {Gratton}, R., \& {Carretta},
  E. 2002, \aj, 124, 470

\bibitem[{{Cohen} {et~al.}(2008){Cohen}, {Christlieb}, {McWilliam}, {Shectman},
  {Thompson}, {Melendez}, {Wisotzki}, \& {Reimers}}]{cohen07-2}
{Cohen}, J.~G., {Christlieb}, N., {McWilliam}, A., {Shectman}, S., {Thompson},
  I., {Melendez}, J., {Wisotzki}, L., \& {Reimers}, D. 2008, \apj, 672, 320

\bibitem[{{Cohen} {et~al.}(2004){Cohen}, {Christlieb}, {McWilliam}, {Shectman},
  {Thompson}, {Wasserburg}, {Ivans}, {Dehn}, {Karlsson}, \&
  {Melendez}}]{cohen04}
{Cohen}, J.~G., {Christlieb}, N., {McWilliam}, A., {Shectman}, S., {Thompson},
  I., {Wasserburg}, G.~J., {Ivans}, I., {Dehn}, M., {Karlsson}, T., \&
  {Melendez}, J. 2004, \apj, 612, 1107

\bibitem[{{Cohen} {et~al.}(2007){Cohen}, {McWilliam}, {Christlieb}, {Shectman},
  {Thompson}, {Melendez}, {Wisotzki}, \& {Reimers}}]{cohen07}
{Cohen}, J.~G., {McWilliam}, A., {Christlieb}, N., {Shectman}, S., {Thompson},
  I., {Melendez}, J., {Wisotzki}, L., \& {Reimers}, D. 2007, \apjl, 659, L161

\bibitem[{{Collet} {et~al.}(2007){Collet}, {Asplund}, \& {Trampedach}}]{collet}
{Collet}, R., {Asplund}, M., \& {Trampedach}, R. 2007, \aap, 469, 687

\bibitem[{{Cowan} {et~al.}(2002){Cowan}, {Sneden}, {Burles}, {Ivans}, {Beers},
  {Truran}, {Lawler}, {Primas}, {Fuller}, {Pfeiffer}, \& {Kratz}}]{cowan02}
{Cowan}, J.~J., {Sneden}, C., {Burles}, S., {Ivans}, I.~I., {Beers}, T.~C.,
  {Truran}, J.~W., {Lawler}, J.~E., {Primas}, F., {Fuller}, G.~M., {Pfeiffer},
  B., \& {Kratz}, K.-L. 2002, \apj, 572, 861

\bibitem[{{Delahaye} \& {Pinsonneault}(2006)}]{delahaye06}
{Delahaye}, F., \& {Pinsonneault}, M.~H. 2006, \apj, 649, 529

\bibitem[{{Fitzpatrick} \& {Sneden}(1987)}]{spectre}
{Fitzpatrick}, M.~J., \& {Sneden}, C. 1987, in Bulletin of the American
  Astronomical Society, Vol.~19, Bulletin of the American Astronomical Society,
  1129--+

\bibitem[{{Fran{\c c}ois} {et~al.}(2007){Fran{\c c}ois}, {Depagne}, {Hill},
  {Spite}, {Spite}, {Plez}, {Beers}, {Andersen}, {James}, {Barbuy}, {Cayrel},
  {Bonifacio}, {Molaro}, {Nordstr{\"o}m}, \& {Primas}}]{francois}
{Fran{\c c}ois}, P., {Depagne}, E., {Hill}, V., {Spite}, M., {Spite}, F.,
  {Plez}, B., {Beers}, T.~C., {Andersen}, J., {James}, G., {Barbuy}, B.,
  {Cayrel}, R., {Bonifacio}, P., {Molaro}, P., {Nordstr{\"o}m}, B., \&
  {Primas}, F. 2007, \aap, 476, 935

\bibitem[{{Frebel} {et~al.}(2005){Frebel}, {Aoki}, {Christlieb}, {Ando},
  {Asplund}, {Barklem}, {Beers}, {Eriksson}, {Fechner}, {Fujimoto}, {Honda},
  {Kajino}, {Minezaki}, {Nomoto}, {Norris}, {Ryan}, {Takada-Hidai},
  {Tsangarides}, \& {Yoshii}}]{frebel05}
{Frebel}, A., {Aoki}, W., {Christlieb}, N., {Ando}, H., {Asplund}, M.,
  {Barklem}, P.~S., {Beers}, T.~C., {Eriksson}, K., {Fechner}, C., {Fujimoto},
  M.~Y., {Honda}, S., {Kajino}, T., {Minezaki}, T., {Nomoto}, K., {Norris},
  J.~E., {Ryan}, S.~G., {Takada-Hidai}, M., {Tsangarides}, S., \& {Yoshii}, Y.
  2005, \nat, 434, 871

\bibitem[{{Frebel} {et~al.}(2007{\natexlab{a}}){Frebel}, {Christlieb},
  {Norris}, {Thom}, {Beers}, \& {Rhee}}]{frebel07}
{Frebel}, A., {Christlieb}, N., {Norris}, J.~E., {Thom}, C., {Beers}, T.~C., \&
  {Rhee}, J. 2007{\natexlab{a}}, \apjl, 660, L117

\bibitem[{{Frebel} {et~al.}(2007{\natexlab{b}}){Frebel}, {Johnson}, \&
  {Bromm}}]{frebel07c}
{Frebel}, A., {Johnson}, J.~L., \& {Bromm}, V. 2007{\natexlab{b}}, \mnras, 380,
  L40

\bibitem[{{Frebel} {et~al.}(2007{\natexlab{c}}){Frebel}, {Norris}, {Aoki},
  {Honda}, {Bessell}, {Takada-Hidai}, {Beers}, \& {Christlieb}}]{frebel07b}
{Frebel}, A., {Norris}, J.~E., {Aoki}, W., {Honda}, S., {Bessell}, M.~S.,
  {Takada-Hidai}, M., {Beers}, T.~C., \& {Christlieb}, N. 2007{\natexlab{c}},
  \apj, 658, 534

\bibitem[{{Fuhr} \& {Wiese}(2006)}]{felines}
{Fuhr}, J.~R., \& {Wiese}, W.~L. 2006, Journal of Physical and Chemical
  Reference Data, 35, 1669

\bibitem[{{Fulbright} {et~al.}(2004){Fulbright}, {Rich}, \&
  {Castro}}]{fulbright04}
{Fulbright}, J.~P., {Rich}, R.~M., \& {Castro}, S. 2004, \apj, 612, 447

\bibitem[{{Gratton} {et~al.}(2000){Gratton}, {Sneden}, {Carretta}, \&
  {Bragaglia}}]{gratton00}
{Gratton}, R.~G., {Sneden}, C., {Carretta}, E., \& {Bragaglia}, A. 2000, \aap,
  354, 169

\bibitem[{{Grevesse} \& {Sauval}(1998)}]{gs98}
{Grevesse}, N., \& {Sauval}, A.~J. 1998, Space Science Reviews, 85, 161

\bibitem[{{Hauck} \& {Mermilliod}(1998)}]{hauck}
{Hauck}, B., \& {Mermilliod}, M. 1998, \aaps, 129, 431

\bibitem[{{Heger} \& {Woosley}(2008)}]{heger08}
{Heger}, A., \& {Woosley}, S.~E. 2008, ApJ, submitted (astro-ph/0803.3161)

\bibitem[{{Hill} {et~al.}(2002){Hill}, {Plez}, {Cayrel}, {Beers},
  {Nordstr{\"o}m}, {Andersen}, {Spite}, {Spite}, {Barbuy}, {Bonifacio},
  {Depagne}, {Fran{\c c}ois}, \& {Primas}}]{hill02}
{Hill}, V., {Plez}, B., {Cayrel}, R., {Beers}, T.~C., {Nordstr{\"o}m}, B.,
  {Andersen}, J., {Spite}, M., {Spite}, F., {Barbuy}, B., {Bonifacio}, P.,
  {Depagne}, E., {Fran{\c c}ois}, P., \& {Primas}, F. 2002, \aap, 387, 560

\bibitem[{{Hoffman} {et~al.}(1996){Hoffman}, {Woosley}, {Fuller}, \&
  {Meyer}}]{hoffman96}
{Hoffman}, R.~D., {Woosley}, S.~E., {Fuller}, G.~M., \& {Meyer}, B.~S. 1996,
  \apj, 460, 478

\bibitem[{{H{\o}g} {et~al.}(2000){H{\o}g}, {Fabricius}, {Makarov}, {Urban},
  {Corbin}, {Wycoff}, {Bastian}, {Schwekendiek}, \& {Wicenec}}]{hog}
{H{\o}g}, E., {Fabricius}, C., {Makarov}, V.~V., {Urban}, S., {Corbin}, T.,
  {Wycoff}, G., {Bastian}, U., {Schwekendiek}, P., \& {Wicenec}, A. 2000, \aap,
  355, L27

\bibitem[{{Honda} {et~al.}(2004){Honda}, {Aoki}, {Kajino}, {Ando}, {Beers},
  {Izumiura}, {Sadakane}, \& {Takada-Hidai}}]{honda04}
{Honda}, S., {Aoki}, W., {Kajino}, T., {Ando}, H., {Beers}, T.~C., {Izumiura},
  H., {Sadakane}, K., \& {Takada-Hidai}, M. 2004, \apj, 607, 474

\bibitem[{{Ivans} {et~al.}(2006){Ivans}, {Simmerer}, {Sneden}, {Lawler},
  {Cowan}, {Gallino}, \& {Bisterzo}}]{ivans06}
{Ivans}, I.~I., {Simmerer}, J., {Sneden}, C., {Lawler}, J.~E., {Cowan}, J.~J.,
  {Gallino}, R., \& {Bisterzo}, S. 2006, \apj, 645, 613

\bibitem[{{Ivans} {et~al.}(2003){Ivans}, {Sneden}, {James}, {Preston},
  {Fulbright}, {H{\"o}flich}, {Carney}, \& {Wheeler}}]{ivans03}
{Ivans}, I.~I., {Sneden}, C., {James}, C.~R., {Preston}, G.~W., {Fulbright},
  J.~P., {H{\"o}flich}, P.~A., {Carney}, B.~W., \& {Wheeler}, J.~C. 2003, \apj,
  592, 906

\bibitem[{{Iwamoto} {et~al.}(2005){Iwamoto}, {Umeda}, {Tominaga}, {Nomoto}, \&
  {Maeda}}]{iwamoto}
{Iwamoto}, N., {Umeda}, H., {Tominaga}, N., {Nomoto}, K., \& {Maeda}, K. 2005,
  Science, 309, 451

\bibitem[{{Johnson}(2002)}]{johnson2002}
{Johnson}, J.~A. 2002, \apjs, 139, 219

\bibitem[{{Johnson} \& {Bolte}(2001)}]{jb01}
{Johnson}, J.~A., \& {Bolte}, M. 2001, \apj, 554, 888

\bibitem[{{Johnson} \& {Bolte}(2002{\natexlab{a}})}]{jbcs22183}
---. 2002{\natexlab{a}}, \apjl, 579, L87

\bibitem[{{Johnson} \& {Bolte}(2002{\natexlab{b}})}]{jb2002}
---. 2002{\natexlab{b}}, \apj, 579, 616

\bibitem[{{Johnson} \& {Bolte}(2004)}]{jbcs31062}
---. 2004, \apj, 605, 462

\bibitem[{{Johnson} {et~al.}(2007){Johnson}, {Herwig}, {Beers}, \&
  {Christlieb}}]{johnson07}
{Johnson}, J.~A., {Herwig}, F., {Beers}, T.~C., \& {Christlieb}, N. 2007, \apj,
  658, 1203

\bibitem[{{Karlsson}(2006)}]{karlsson}
{Karlsson}, T. 2006, \apjl, 641, L41

\bibitem[{{Kim} {et~al.}(2002){Kim}, {Demarque}, {Yi}, \&
  {Alexander}}]{kim2002}
{Kim}, Y.-C., {Demarque}, P., {Yi}, S.~K., \& {Alexander}, D.~R. 2002, \apjs,
  143, 499

\bibitem[{{Kraft}(1994)}]{kraft94}
{Kraft}, R.~P. 1994, \pasp, 106, 553

\bibitem[{{Kratz} {et~al.}(2007){Kratz}, {Farouqi}, {Pfeiffer}, {Truran},
  {Sneden}, \& {Cowan}}]{kratz07}
{Kratz}, K.-L., {Farouqi}, K., {Pfeiffer}, B., {Truran}, J.~W., {Sneden}, C.,
  \& {Cowan}, J.~J. 2007, \apj, 662, 39

\bibitem[{{Lai} {et~al.}(2004){Lai}, {Bolte}, {Johnson}, \&
  {Lucatello}}]{lai04}
{Lai}, D.~K., {Bolte}, M., {Johnson}, J.~A., \& {Lucatello}, S. 2004, \aj, 128,
  2402

\bibitem[{{Ljung} {et~al.}(2006){Ljung}, {Nilsson}, {Asplund}, \&
  {Johansson}}]{ljung}
{Ljung}, G., {Nilsson}, H., {Asplund}, M., \& {Johansson}, S. 2006, \aap, 456,
  1181

\bibitem[{{M{\aa}rtensson-Pendrill} {et~al.}(1994){M{\aa}rtensson-Pendrill},
  {Gough}, \& {Hannaford}}]{ybhfs}
{M{\aa}rtensson-Pendrill}, A.-M., {Gough}, D.~S., \& {Hannaford}, P. 1994,
  \pra, 49, 3351

\bibitem[{{Malcheva} {et~al.}(2006){Malcheva}, {Blagoev}, {Mayo}, {Ortiz},
  {Xu}, {Svanberg}, {Quinet}, \& {Bi{\'e}mont}}]{malcheva}
{Malcheva}, G., {Blagoev}, K., {Mayo}, R., {Ortiz}, M., {Xu}, H.~L.,
  {Svanberg}, S., {Quinet}, P., \& {Bi{\'e}mont}, E. 2006, \mnras, 367, 754

\bibitem[{{Masseron} {et~al.}(2006){Masseron}, {van Eck}, {Famaey}, {Goriely},
  {Plez}, {Siess}, {Beers}, {Primas}, \& {Jorissen}}]{masseron06}
{Masseron}, T., {van Eck}, S., {Famaey}, B., {Goriely}, S., {Plez}, B.,
  {Siess}, L., {Beers}, T.~C., {Primas}, F., \& {Jorissen}, A. 2006, \aap, 455,
  1059

\bibitem[{{McWilliam}(1998)}]{mcwilliam98}
{McWilliam}, A. 1998, \aj, 115, 1640

\bibitem[{{McWilliam} {et~al.}(1995){McWilliam}, {Preston}, {Sneden}, \&
  {Searle}}]{mcwilliam95}
{McWilliam}, A., {Preston}, G.~W., {Sneden}, C., \& {Searle}, L. 1995, \aj,
  109, 2757

\bibitem[{{Mishenina} {et~al.}(2002){Mishenina}, {Kovtyukh}, {Soubiran},
  {Travaglio}, \& {Busso}}]{mishenina02}
{Mishenina}, T.~V., {Kovtyukh}, V.~V., {Soubiran}, C., {Travaglio}, C., \&
  {Busso}, M. 2002, \aap, 396, 189

\bibitem[{{Montes} {et~al.}(2007){Montes}, {Beers}, {Cowan}, {Elliot},
  {Farouqi}, {Gallino}, {Heil}, {Kratz}, {Pfeiffer}, {Pignatari}, \&
  {Schatz}}]{montes}
{Montes}, F., {Beers}, T.~C., {Cowan}, J., {Elliot}, T., {Farouqi}, K.,
  {Gallino}, R., {Heil}, M., {Kratz}, K.~., {Pfeiffer}, B., {Pignatari}, M., \&
  {Schatz}, H. 2007, ArXiv e-prints, 709

\bibitem[{{Nilsson} {et~al.}(2006){Nilsson}, {Ljung}, {Lundberg}, \&
  {Nielsen}}]{crIIlines}
{Nilsson}, H., {Ljung}, G., {Lundberg}, H., \& {Nielsen}, K.~E. 2006, \aap,
  445, 1165

\bibitem[{{Nilsson} {et~al.}(2002){Nilsson}, {Zhang}, {Lundberg}, {Johansson},
  \& {Nordstr{\"o}m}}]{nilsson02}
{Nilsson}, H., {Zhang}, Z.~G., {Lundberg}, H., {Johansson}, S., \&
  {Nordstr{\"o}m}, B. 2002, \aap, 382, 368

\bibitem[{{Norris} {et~al.}(2007){Norris}, {Christlieb}, {Korn}, {Eriksson},
  {Bessell}, {Beers}, {Wisotzki}, \& {Reimers}}]{norris07}
{Norris}, J.~E., {Christlieb}, N., {Korn}, A.~J., {Eriksson}, K., {Bessell},
  M.~S., {Beers}, T.~C., {Wisotzki}, L., \& {Reimers}, D. 2007, \apj, 670, 774

\bibitem[{{Norris} {et~al.}(1999){Norris}, {Ryan}, \& {Beers}}]{norris99}
{Norris}, J.~E., {Ryan}, S.~G., \& {Beers}, T.~C. 1999, \apjs, 123, 639

\bibitem[{{Pinto} \& {Woosley}(1988)}]{pinto88}
{Pinto}, P.~A., \& {Woosley}, S.~E. 1988, \nat, 333, 534

\bibitem[{{Preston} {et~al.}(2006){Preston}, {Sneden}, {Thompson}, {Shectman},
  \& {Burley}}]{preston06}
{Preston}, G.~W., {Sneden}, C., {Thompson}, I.~B., {Shectman}, S.~A., \&
  {Burley}, G.~S. 2006, \aj, 132, 85

\bibitem[{{Prochaska} \& {McWilliam}(2000)}]{prochaska00}
{Prochaska}, J.~X., \& {McWilliam}, A. 2000, \apjl, 537, L57

\bibitem[{{Pruet} {et~al.}(2005){Pruet}, {Woosley}, {Buras}, {Janka}, \&
  {Hoffman}}]{pruet}
{Pruet}, J., {Woosley}, S.~E., {Buras}, R., {Janka}, H.-T., \& {Hoffman}, R.~D.
  2005, \apj, 623, 325

\bibitem[{{Qian}(2002)}]{qian2002}
{Qian}, Y.-Z. 2002, \apjl, 569, L103

\bibitem[{{Qian} \& {Wasserburg}(2001)}]{qian2001}
{Qian}, Y.-Z., \& {Wasserburg}, G.~J. 2001, \apj, 559, 925

\bibitem[{{Qian} \& {Wasserburg}(2007)}]{qian2007}
---. 2007, \physrep, 442, 237

\bibitem[{{Ram{\'{\i}}rez} \& {Mel{\'e}ndez}(2005)}]{ramirez05}
{Ram{\'{\i}}rez}, I., \& {Mel{\'e}ndez}, J. 2005, \apj, 626, 465

\bibitem[{{Romano} \& {Matteucci}(2007)}]{romano}
{Romano}, D., \& {Matteucci}, F. 2007, \mnras, 378, L59

\bibitem[{{Ryan} {et~al.}(1996){Ryan}, {Norris}, \& {Beers}}]{ryan96}
{Ryan}, S.~G., {Norris}, J.~E., \& {Beers}, T.~C. 1996, \apj, 471, 254

\bibitem[{{Salvadori} {et~al.}(2007){Salvadori}, {Schneider}, \&
  {Ferrara}}]{salvadori}
{Salvadori}, S., {Schneider}, R., \& {Ferrara}, A. 2007, \mnras, 381, 647

\bibitem[{{Schlegel} {et~al.}(1998){Schlegel}, {Finkbeiner}, \&
  {Davis}}]{schlegel}
{Schlegel}, D.~J., {Finkbeiner}, D.~P., \& {Davis}, M. 1998, \apj, 500, 525

\bibitem[{{Schuster} {et~al.}(2004){Schuster}, {Beers}, {Michel}, {Nissen}, \&
  {Garc{\'{\i}}a}}]{schuster04}
{Schuster}, W.~J., {Beers}, T.~C., {Michel}, R., {Nissen}, P.~E., \&
  {Garc{\'{\i}}a}, G. 2004, \aap, 422, 527

\bibitem[{{Simmerer} {et~al.}(2004){Simmerer}, {Sneden}, {Cowan}, {Collier},
  {Woolf}, \& {Lawler}}]{simmerer04}
{Simmerer}, J., {Sneden}, C., {Cowan}, J.~J., {Collier}, J., {Woolf}, V.~M., \&
  {Lawler}, J.~E. 2004, \apj, 617, 1091

\bibitem[{{Simmerer} {et~al.}(2003){Simmerer}, {Sneden}, {Ivans}, {Kraft},
  {Shetrone}, \& {Smith}}]{simmerer03}
{Simmerer}, J., {Sneden}, C., {Ivans}, I.~I., {Kraft}, R.~P., {Shetrone},
  M.~D., \& {Smith}, V.~V. 2003, \aj, 125, 2018

\bibitem[{{Sneden} {et~al.}(2003){Sneden}, {Cowan}, {Lawler}, {Ivans},
  {Burles}, {Beers}, {Primas}, {Hill}, {Truran}, {Fuller}, {Pfeiffer}, \&
  {Kratz}}]{sneden03}
{Sneden}, C., {Cowan}, J.~J., {Lawler}, J.~E., {Ivans}, I.~I., {Burles}, S.,
  {Beers}, T.~C., {Primas}, F., {Hill}, V., {Truran}, J.~W., {Fuller}, G.~M.,
  {Pfeiffer}, B., \& {Kratz}, K.-L. 2003, \apj, 591, 936

\bibitem[{{Sneden} \& {Crocker}(1988)}]{sneden88}
{Sneden}, C., \& {Crocker}, D.~A. 1988, \apj, 335, 406

\bibitem[{{Sneden} {et~al.}(1996){Sneden}, {McWilliam}, {Preston}, {Cowan},
  {Burris}, \& {Armosky}}]{sneden96}
{Sneden}, C., {McWilliam}, A., {Preston}, G.~W., {Cowan}, J.~J., {Burris},
  D.~L., \& {Armosky}, B.~J. 1996, \apj, 467, 819

\bibitem[{{Sobeck} {et~al.}(2007){Sobeck}, {Lawler}, \& {Sneden}}]{sobeck07}
{Sobeck}, J.~S., {Lawler}, J.~E., \& {Sneden}, C. 2007, \apj, 667, 1267

\bibitem[{{Spite} {et~al.}(2006){Spite}, {Cayrel}, {Hill}, {Spite}, {Fran{\c
  c}ois}, {Plez}, {Bonifacio}, {Molaro}, {Depagne}, {Andersen}, {Barbuy},
  {Beers}, {Nordstr{\"o}m}, \& {Primas}}]{spite07}
{Spite}, M., {Cayrel}, R., {Hill}, V., {Spite}, F., {Fran{\c c}ois}, P.,
  {Plez}, B., {Bonifacio}, P., {Molaro}, P., {Depagne}, E., {Andersen}, J.,
  {Barbuy}, B., {Beers}, T.~C., {Nordstr{\"o}m}, B., \& {Primas}, F. 2006,
  \aap, 455, 291

\bibitem[{{Spite} {et~al.}(2005){Spite}, {Cayrel}, {Plez}, {Hill}, {Spite},
  {Depagne}, {Fran{\c c}ois}, {Bonifacio}, {Barbuy}, {Beers}, {Andersen},
  {Molaro}, {Nordstr{\"o}m}, \& {Primas}}]{spite05}
{Spite}, M., {Cayrel}, R., {Plez}, B., {Hill}, V., {Spite}, F., {Depagne}, E.,
  {Fran{\c c}ois}, P., {Bonifacio}, P., {Barbuy}, B., {Beers}, T., {Andersen},
  J., {Molaro}, P., {Nordstr{\"o}m}, B., \& {Primas}, F. 2005, \aap, 430, 655

\bibitem[{{Suda} {et~al.}(2004){Suda}, {Aikawa}, {Machida}, {Fujimoto}, \&
  {Iben}}]{suda04}
{Suda}, T., {Aikawa}, M., {Machida}, M.~N., {Fujimoto}, M.~Y., \& {Iben}, I.~J.
  2004, \apj, 611, 476

\bibitem[{{Travaglio} {et~al.}(2004){Travaglio}, {Gallino}, {Arnone}, {Cowan},
  {Jordan}, \& {Sneden}}]{travaglio}
{Travaglio}, C., {Gallino}, R., {Arnone}, E., {Cowan}, J., {Jordan}, F., \&
  {Sneden}, C. 2004, \apj, 601, 864

\bibitem[{{Tsujimoto} \& {Shigeyama}(2003)}]{tsuj2003}
{Tsujimoto}, T., \& {Shigeyama}, T. 2003, \apjl, 584, L87

\bibitem[{{Tumlinson}(2006{\natexlab{a}})}]{tumlinson06-2}
{Tumlinson}, J. 2006{\natexlab{a}}, \apj, 641, 1

\bibitem[{{Tumlinson}(2006{\natexlab{b}})}]{tumlinson06}
---. 2006{\natexlab{b}}, New Astronomy Review, 50, 101

\bibitem[{{Venn} \& {Lambert}(2008)}]{venn08}
{Venn}, K.~A., \& {Lambert}, D.~L. 2008, \apj, 677, 572

\bibitem[{{Vogt} {et~al.}(1994){Vogt}, {Allen}, {Bigelow}, {Bresee}, {Brown},
  {Cantrall}, {Conrad}, {Couture}, {Delaney}, {Epps}, {Hilyard}, {Hilyard},
  {Horn}, {Jern}, {Kanto}, {Keane}, {Kibrick}, {Lewis}, {Osborne},
  {Pardeilhan}, {Pfister}, {Ricketts}, {Robinson}, {Stover}, {Tucker}, {Ward},
  \& {Wei}}]{vogt94}
{Vogt}, S.~S., {Allen}, S.~L., {Bigelow}, B.~C., {Bresee}, L., {Brown}, B.,
  {Cantrall}, T., {Conrad}, A., {Couture}, M., {Delaney}, C., {Epps}, H.~W.,
  {Hilyard}, D., {Hilyard}, D.~F., {Horn}, E., {Jern}, N., {Kanto}, D.,
  {Keane}, M.~J., {Kibrick}, R.~I., {Lewis}, J.~W., {Osborne}, J.,
  {Pardeilhan}, G.~H., {Pfister}, T., {Ricketts}, T., {Robinson}, L.~B.,
  {Stover}, R.~J., {Tucker}, D., {Ward}, J., \& {Wei}, M.~Z. 1994, in Presented
  at the Society of Photo-Optical Instrumentation Engineers (SPIE) Conference,
  Vol. 2198, Proc. SPIE Instrumentation in Astronomy VIII, David L. Crawford;
  Eric R. Craine; Eds., Volume 2198, p. 362, ed. D.~L. {Crawford} \& E.~R.
  {Craine}, 362--+

\bibitem[{{Westin} {et~al.}(2000){Westin}, {Sneden}, {Gustafsson}, \&
  {Cowan}}]{westin}
{Westin}, J., {Sneden}, C., {Gustafsson}, B., \& {Cowan}, J.~J. 2000, \apj,
  530, 783

\bibitem[{{Woosley} \& {Weaver}(1995)}]{ww95}
{Woosley}, S.~E., \& {Weaver}, T.~A. 1995, \apjs, 101, 181

\bibitem[{{Zhang} {et~al.}(2008){Zhang}, {Woosley}, \& {Heger}}]{zhang}
{Zhang}, W., {Woosley}, S.~E., \& {Heger}, A. 2008, \apj, 679, 639

\end{thebibliography}
 
\clearpage

\begin{deluxetable}{lrrrlcrrrr}
\tablecolumns{10}
\tabletypesize{\scriptsize}
\tablewidth{0pc}
\tablecaption{Observation details \label{odetails}}
\tablehead{
\colhead{Star} & \colhead{$V_{mag}$}   & \colhead{V Ref.} &
\colhead{$v_r$} & \colhead{Observation}
&\colhead{Total Exposure}  & \colhead{Wavelength} &\colhead{S/N} &\colhead{S/N} &\colhead{S/N} \\ \colhead{ID} &
\colhead{} & \colhead{} &  \colhead{(km s$^{-1}$)\tablenotemark{*}} & \colhead{Date (UT)} & \colhead{Time (s)} &
\colhead{Coverage (\AA)} & \colhead{5560 \AA} &\colhead{4500 \AA} &\colhead{3435 \AA} }
\startdata
BD+03 740     &   9.80    &  1  &   173.1 (0.2) & 2004, Oct 7-8   &  3900    &      3020-5870  &   425      &  340     &  278     \\
BD+23 3130    &   8.94    &  2  &$-$285.1 (0.2) & 2006, Aug 19    &  960     &      3055-5893  &   424      &  245     &  161     \\
BD+24 1676    &   10.79   &  1  &$-$237.8 (0.2) & 2004, Oct 7     &  3600    &      3020-5870  &   320      &  288     &  193     \\
BS 16077-007  &   12.41   &  3  & $-$38.3 (0.2) & 2005, May 29-30 &  2400    &      3020-5870  &   142      &  111     &  61      \\
BS 16080-054  &   12.776  &  4  &$-$121.2 (0.2) & 2004, Apr 7     &  1200    &      3220-4655  &   \nodata  &  78      &  12      \\
              &           &     &$-$119.3 (0.3) & 2004, Apr 9     &  1200    &      5385-7665  &   130      &  \nodata &  \nodata \\
BS 16080-093  &   13.61   &  5  &$-$205.2 (0.2) & 2001, Aug 25    &  5400    &      3770-5280  &   \nodata  &  85      &  \nodata \\
BS 16084-160  &   13.14   &  3  &$-$123.2 (0.2) & 2001, Aug 25    &  7200    &      3770-5280  &   \nodata  &  103     &  \nodata \\
\enddata
\tablenotetext{*}{Heliocentric velocity and estimated internal error.}
\tablerefs{(1) \citet{hauck}; (2) \citet{hog}; (3)
  \citet{schuster04}; (4) \citet{aoki05}; (5) \citet{bonifacio00};
  (6) \citet{cohen04}; (7) \citet{norris99}; (8) \citet{honda04}; (9) \citet{beers00}}
\end{deluxetable}
\clearpage

\begin{landscape}
\begin{deluxetable}{llrlcccccccccc}
\tablecolumns{14}
\setlength{\tabcolsep}{0.04in}
\tabletypesize{\tiny}
\tablewidth{0pc}
\tablecaption{Equivalent Widths and Atomic Parameters (full version
  available in online edition) \label{ew1}}
\tablehead{
  \colhead{Wavelength} & \colhead{Element} & \colhead{log$gf$} & \colhead{EP} & \colhead{BD+03 740} & \colhead{BD+23 3130} & \colhead{BD+24 1670} & \colhead{BS 16077-007} & \colhead{BS 16080-054
} & \colhead{BS 16080-093} & \colhead{BS 16084-160} & \colhead{BS 16545-089} & \colhead{BS 16550-087} & \colhead{BS 16928-053}}
\startdata
  5889.95 &   11.0 &    0.11 &    0.00 &     \nodata &      128.50 &     \nodata &     \nodata &      173.50 &     \nodata &      124.40 &       25.00 &      121.00 &      145.70   \\ 
  5895.92 &   11.0 & $-$0.19 &    0.00 &     \nodata &      107.50 &     \nodata &     \nodata &      141.00 &     \nodata &       97.70 &     \nodata &       94.50 &      123.60   \\ 
  3829.36 &   12.0 & $-$0.21 &    2.71 &     \nodata &      160.70 &     \nodata &     \nodata &      152.40 &      140.40 &      136.40 &       60.40 &      139.10 &      144.20   \\ 
  3832.31 &   12.0 &    0.15 &    2.71 &     \nodata &     \nodata &     \nodata &     \nodata &      187.30 &      167.20 &      157.70 &     \nodata &      167.10 &      180.20   \\ 
  4057.52 &   12.0 & $-$0.90 &    4.34 &        6.70 &       28.60 &       14.30 &     \nodata &     \nodata &       15.50 &     \nodata &        2.00 &       13.20 &     \nodata   
\enddata
\end{deluxetable}
\clearpage
\end{landscape}

\begin{landscape}
\begin{deluxetable}{llrlccccccccc}
\tablecolumns{13}
\setlength{\tabcolsep}{0.04in}
\tabletypesize{\tiny}
\tablewidth{0pc}
\tablecaption{Equivalent Widths and Atomic Parameters (full version
  available in online edition) \label{ew2}}
\tablehead{
  \colhead{Wavelength} & \colhead{Element} & \colhead{log$gf$} & \colhead{EP} & \colhead{BS 16929-005} & \colhead{BS 16467-062} & \colhead{CS 22872-102} & \colhead{CS 22878-027} & \colhead{CS 22880-086} & \colhead{CS 22884-108} & \colhead{CS 22944-032} & \colhead{CS 22957-022} & \colhead{CS 22963-004}}
\startdata
  5889.95 &   11.0 &    0.11 &    0.00 &       89.60 &     \nodata &       79.50 &       83.20 &     \nodata &     \nodata &     \nodata &     \nodata &     \nodata   \\ 
  5895.92 &   11.0 & $-$0.19 &    0.00 &       73.30 &     \nodata &       63.90 &       50.00 &     \nodata &     \nodata &     \nodata &     \nodata &     \nodata   \\ 
  3829.36 &   12.0 & $-$0.21 &    2.71 &      118.30 &       80.00 &     \nodata &     \nodata &     \nodata &       79.50 &      137.90 &      133.10 &       96.90   \\ 
  3832.31 &   12.0 &    0.15 &    2.71 &     \nodata &     \nodata &     \nodata &     \nodata &     \nodata &     \nodata &     \nodata &     \nodata &     \nodata   \\ 
  4057.52 &   12.0 & $-$0.90 &    4.34 &       11.60 &     \nodata &        9.60 &     \nodata &     \nodata &     \nodata &       17.10 &       19.90 &     \nodata  
\enddata
\end{deluxetable}
\clearpage
\end{landscape}

\begin{landscape}
\begin{deluxetable}{llrlccccccccc}
\tablecolumns{13}
\setlength{\tabcolsep}{0.04in}
\tabletypesize{\tiny}
\tablewidth{0pc}
\tablecaption{Equivalent Widths and Atomic Parameters (full version
  available in online version)\label{ew3}}
\tablehead{
  \colhead{Wavelength} & \colhead{Element} & \colhead{log$gf$} & \colhead{EP} & \colhead{CS 22965-054} & \colhead{CS 29502-092} & \colhead{CS 29506-007} & \colhead{CS 29522-046} & \colhead{CS 30312-059} & \colhead{CS 30325-028} & \colhead{CS 30336-049} & \colhead{CS 31078-018} & \colhead{CS 31085-024}}
\startdata
  5889.95 &   11.0 &    0.11 &    0.00 &     \nodata &     \nodata &     \nodata &     \nodata &      126.90 &     \nodata &     \nodata &     \nodata &     \nodata   \\ 
  5895.92 &   11.0 & $-$0.19 &    0.00 &     \nodata &     \nodata &     \nodata &     \nodata &      104.90 &     \nodata &     \nodata &     \nodata &     \nodata   \\ 
  3829.36 &   12.0 & $-$0.21 &    2.71 &       95.40 &      136.70 &     \nodata &     \nodata &      135.90 &      157.20 &       83.00 &      152.00 &      138.00   \\ 
  3832.31 &   12.0 &    0.15 &    2.71 &     \nodata &     \nodata &     \nodata &     \nodata &     \nodata &     \nodata &      103.70 &     \nodata &     \nodata   \\ 
  4057.52 &   12.0 & $-$0.90 &    4.34 &     \nodata &       18.80 &     \nodata &       46.70 &       16.40 &       27.70 &     \nodata &       20.90 &       11.10  
\enddata
\end{deluxetable}
\clearpage
\end{landscape}

\begin{landscape}
\begin{deluxetable}{lrrrrrrrrrrrrr}
\tablecolumns{14}
\tabletypesize{\scriptsize}
\tablewidth{0pc}
\tablecaption{Stellar Parameters \label{atm}}
\tablehead{
\colhead{} & \colhead{} & \colhead{} & \colhead{} & \colhead{} & \multicolumn{3}{c}{All lines} &
\multicolumn{3}{c}{$\chi > 0.2$eV} & \multicolumn{3}{c}{$\chi > 1.2$eV}
\\\cline{6-14} 
\\ \colhead{Star} & \colhead{[Fe/H]} & \colhead{\teff{}} & \colhead{log$g$} & \colhead{v$_t$} 
& \colhead{$\chi$ trend} & \colhead{r} &\colhead{Numb.} & \colhead{$\chi$ trend} & \colhead{r} 
&\colhead{Numb.} & \colhead{$\chi$ trend} & \colhead{r} &\colhead{Numb.}
\\ \colhead{ID} & \colhead{} & \colhead{(K)} & \colhead{(cgs)} & \colhead{(km s$^{-1}$)} 
& \colhead{(dex/eV)} & \colhead{} & \colhead{lines} & \colhead{(dex/eV)} & \colhead{} & 
\colhead{lines} & \colhead{(dex/eV)} & \colhead{} & \colhead{lines} }
\startdata    
    BD+03 740  & $-$2.75 & 6557 & 4.28 & 1.8 & $-$0.04 & $-$0.49 & 102 & $-$0.03 & $-$0.38 &  91 & $-$0.03 & $-$0.27 &  76  \\
   BD+23 3130  & $-$2.60 & 5285 & 2.83 & 1.6 & $-$0.07 & $-$0.64 & 150 & $-$0.07 & $-$0.59 & 133 & $-$0.04 & $-$0.36 & 107  \\
   BD+24 1676  & $-$2.60 & 6241 & 3.81 & 1.6 & $-$0.01 & $-$0.21 & 119 & $-$0.01 & $-$0.14 & 107 &    0.00 & $-$0.02 &  88  \\
  BS 16077-007  & $-$2.80 & 6544 & 4.29 & 1.6 & $-$0.05 & $-$0.56 &  77 & $-$0.06 & $-$0.54 &  67 & $-$0.03 & $-$0.29 &  52  \\
  BS 16080-054  & $-$3.00 & 4805 & 1.48 & 2.1 & $-$0.03 & $-$0.29 &  91 & $-$0.02 & $-$0.05 &  80 & $-$0.01 & $-$0.05 &  63  \\
  BS 16080-093  & $-$3.20 & 4940 & 1.86 & 1.9 & $-$0.09 & $-$0.69 &  94 & $-$0.07 & $-$0.52 &  78 & $-$0.04 & $-$0.23 &  61  \\
  BS 16084-160  & $-$3.15 & 4728 & 1.26 & 2.1 & $-$0.09 & $-$0.59 & 155 & $-$0.07 & $-$0.45 & 137 & $-$0.03 & $-$0.17 & 105  \\
  BS 16545-089  & $-$3.50 & 6500 & 4.25 & 1.6 & $-$0.03 & $-$0.30 &  27 & $-$0.01 & $-$0.03 &  21 & $-$0.10 & $-$0.50 &  16  \\
\enddata
\end{deluxetable}
\clearpage
\end{landscape}

\begin{deluxetable}{lrrrrrrrr}
\tablecolumns{9}
\tabletypesize{\small}
\tablecaption{Comparison of Atmospheric Parameters \label{atmcomp}}
\tablewidth{0pc}
\tablehead{
\colhead{}  & \multicolumn{3}{c}{This Study} & \colhead{} &\multicolumn{3}{c}{Previous Study} & \colhead{}\\
\cline{2-4}  \cline{6-8} \\
\colhead{Star ID} & \colhead{\teff{}} & \colhead{\logg{}} & \colhead{v$_t$} & \colhead{} & \colhead{\teff{}} & \colhead{\logg{}} 
& \colhead{v$_t$} & \colhead{Ref.} \\
}
\startdata
BD+03 740    & 6557 & 4.28 & 1.8 & \phn & 6330 & 3.55 & 1.4 & 1\\
BD+24 1676   & 6241 & 3.81 & 1.6 & \phn & 6250 & 3.45 & 1.2 & 1\\
BS 16080-054 & 4805 & 1.48 & 2.1 & \phn & 4800 & 1.1 & 2.4 & 2\\
BS 16084-160 & 4728 & 1.26 & 2.1 & \phn & 4650 & 1.1 & 2.2 & 2\\
CS 30312-059 & 5021 & 2.06 & 1.8 & \phn & 4950 & 2.0 & 1.8 & 2\\
CS 30325-028 & 4911 & 1.70 & 1.8 & \phn & 4900 & 1.8 & 2.0 & 2\\
BS 16467-062 & 5388 & 3.04 & 1.7 & \phn & 5200 & 2.5 & 1.6 & 3\\
BS 16467-062 & 5388 & 3.04 & 1.7 & \phn & 5364 & 2.95 & 1.6 & 4
\enddata
\tablerefs{(1) \citet{ivans03}; (2) \citet{aoki05}; (3) \citet{cayrel04}; (4) \citet{cohen07-2}}
\end{deluxetable}


\begin{deluxetable}{lrrrrr}
\tablecolumns{6}
\tabletypesize{\scriptsize}
\tablewidth{0pc}
\tablecaption{Atmospheric errors for BD+03 740 \label{erBD+03}}
\tablehead{
\colhead{Element} & \colhead{\teff{}} & \colhead{\logg{}} & \colhead{v$_t$} & \colhead{$\sigma_{\mbox{log}\epsilon}$}
 & \colhead{$\sigma_{[X/Fe]}$}\\ 
\colhead{\phn} & \colhead{+100 K} & \colhead{+0.2 dex} & \colhead{+0.2 km s$^{-1}$} & \colhead{} & \colhead{}}
\startdata    
  Fe {\scriptsize I}  &   0.08  &   0.00  &$-$0.02  &   0.08  &   0.00  \\
  Fe {\scriptsize II}  &   0.02  &   0.07  &   0.00  &   0.05  &   0.00  \\
   Li\tablenotemark{1}  &   0.07  &   0.00  &   0.07  &   0.10  &   0.09  \\
    C  &   0.13  &   0.00  &$-$0.05  &   0.14  &   0.06  \\
  Na\tablenotemark{1}   &   0.07  &$-$0.01  &   0.07  &   0.11  &   0.09  \\
  Mg   &   0.05  &$-$0.02  &$-$0.02  &   0.07  &   0.01  \\
  Al   &   0.07  &   0.00  &$-$0.01  &   0.07  &   0.01  \\
  Si   &   0.07  &$-$0.02  &$-$0.06  &   0.10  &   0.03  \\
\enddata
\tablenotetext{1}{EWs from CS 22878-027}
\end{deluxetable}

\clearpage

\begin{deluxetable}{lrrrrr}
\tablecolumns{6}
\tabletypesize{\scriptsize}
\tablewidth{0pc}
\tablecaption{Atmospheric errors for CS 31078-018 \label{erCS31078}}
\tablehead{
\colhead{Element} & \colhead{\teff{}} & \colhead{\logg{}} & \colhead{v$_t$} & \colhead{$\sigma_{\mbox{log}\epsilon}$}
 & \colhead{$\sigma_{[X/Fe]}$}\\ 
\colhead{} & \colhead{+100 K} & \colhead{+0.2 dex} & \colhead{+0.2 km s$^{-1}$} & \colhead{} & \colhead{}}
\startdata    
  Fe {\scriptsize I}  &   0.10  &$-$0.02  &$-$0.05  &   0.03  &   0.00  \\
  Fe {\scriptsize II}  &   0.02  &   0.07  &$-$0.02  &   0.10  &   0.00  \\
    C  &   0.20  &$-$0.05  &   0.00  &   0.12  &   0.11  \\
    N  &   0.20  &$-$0.05  &   0.00  &   0.12  &   0.11  \\
    O  &   0.20  &$-$0.10  &   0.00  &   0.09  &   0.14  \\
  Na\tablenotemark{1}   &   0.13  &$-$0.05  &$-$0.07  &   0.08  &   0.11  \\
  Mg   &   0.09  &$-$0.04  &$-$0.02  &   0.05  &   0.09  \\
  Al   &   0.11  &$-$0.03  &$-$0.07  &   0.04  &   0.05  \\
\enddata
\tablenotetext{1}{EW from BD+23 3130}
\tablenotetext{2}{EW from BS 16929-005}
\end{deluxetable}

\clearpage
\begin{deluxetable}{lrrrrr}
\tablecolumns{6}
\tabletypesize{\scriptsize}
\tablewidth{0pc}
\tablecaption{Atmospheric errors for CS 29502-092 \label{erCS29502}}
\tablehead{
\colhead{Element} & \colhead{\teff{}} & \colhead{\logg{}} & \colhead{v$_t$} & \colhead{$\sigma_{\mbox{log}\epsilon}$}
 & \colhead{$\sigma_{[X/Fe]}$}\\ 
\colhead{} & \colhead{+100 K} & \colhead{+0.2 dex} & \colhead{+0.2 km s$^{-1}$} & \colhead{} & \colhead{}}
\startdata    
  Fe {\scriptsize I}  &   0.12  &$-$0.03  &$-$0.06  &   0.09  &   0.00  \\
  Fe {\scriptsize II}  &   0.01  &   0.06  &$-$0.02  &   0.08  &   0.00  \\
    C  &   0.25  &$-$0.10  &   0.00  &   0.09  &   0.08  \\
    N  &   0.30  &   0.00  &   0.00  &   0.30  &   0.22  \\
    O  &   0.25  &$-$0.10  &   0.00  &   0.09  &   0.08  \\
  Mg   &   0.11  &$-$0.05  &$-$0.04  &   0.01  &   0.04  \\
  Na\tablenotemark{1}   &   0.14  &$-$0.08  &$-$0.09  &   0.05  &   0.06  \\
  Al   &   0.12  &$-$0.04  &$-$0.09  &   0.09  &   0.03  \\
\enddata
\tablenotetext{1}{EWs from BS 16080-054}
\tablenotetext{2}{EWs from CS 30325-028}
\end{deluxetable}
\clearpage

\begin{landscape}
\begin{deluxetable}{lrrrrrrrrrrrrrrrrrrrr}
\tablecolumns{21}
\setlength{\tabcolsep}{0.055in}
\tabletypesize{\tiny}
\tablewidth{0pc}
\tablecaption{Abundances relative to Fe (and \ion{Fe}{2} for ionized species) \label{abund1}}
\tablehead{
   \colhead{Star}  & \colhead{  [Fe/H]} & \colhead{numb.} & \colhead{$\sigma$} & \colhead{total}  
   & \colhead{[Fe {\tiny II}/H]} & \colhead{numb.} & \colhead{$\sigma$} & \colhead{total}  
   & \colhead{  [C/Fe]} & \colhead{numb.} & \colhead{$\sigma$} & \colhead{total}  & \colhead{  [N/Fe]} 
   & \colhead{numb.} & \colhead{$\sigma$} & \colhead{total}  & \colhead{  [O/Fe]} & \colhead{numb.} & \colhead{$\sigma$} &
\colhead{total}
\\
\colhead{ID} & \colhead{} & \colhead{lines} &\colhead{lines} & \colhead{error} & \colhead{} & \colhead{lines} &\colhead{lines} 
& \colhead{error} & \colhead{} & \colhead{lines} &\colhead{lines} & \colhead{error} & \colhead{} & \colhead{lines} &\colhead{lines} 
& \colhead{error} & \colhead{} & \colhead{lines} &\colhead{lines} & \colhead{error}
}
\startdata
   BD+03 740  & $-$2.71 & 102 &    0.10 &    0.08 & $-$2.69 &  10 &    0.10 &    0.06 &    0.59 &   1 &    0.30 &    0.31 & $<$0.29 &   1 & \nodata & \nodata & $<$0.98 &   1 & \nodata & \nodata  \\
  BD+23 3130  & $-$2.58 & 150 &    0.13 &    0.03 & $-$2.62 &  21 &    0.16 &    0.11 &    0.11 &   1 &    0.20 &    0.23 & $-$0.54 &   1 &    0.20 &    0.23 &    0.55 &   1 &    0.30 &    0.33  \\
  BD+24 1676  & $-$2.63 & 119 &    0.07 &    0.08 & $-$2.60 &  14 &    0.09 &    0.06 &    0.41 &   1 &    0.20 &    0.21 & $<$0.21 &   1 & \nodata & \nodata &    0.65 &   1 &    0.30 & \nodata  \\
 BS 16077-007 & $-$2.72 &  77 &    0.10 &    0.08 & $-$2.82 &   4 &    0.14 &    0.09 &    0.60 &   1 &    0.30 &    0.31 & $<$0.80 &   1 & \nodata & \nodata & $<$1.39 &   1 & \nodata & \nodata  \\
 BS 16080-054 & $-$3.07 &  91 &    0.12 &    0.09 & $-$2.94 &  11 &    0.10 &    0.09 & $-$0.45 &   1 &    0.20 &    0.22 &    0.75 &   1 &    0.30 &    0.37 & \nodata &   0 & \nodata & \nodata  \\
 BS 16080-093 & $-$3.19 &  94 &    0.14 &    0.09 & $-$3.11 &  12 &    0.14 &    0.09 &$<-$0.63 &   1 & \nodata & \nodata & \nodata &   0 & \nodata & \nodata & \nodata &   0 & \nodata & \nodata  \\
 BS 16084-160 & $-$3.20 & 155 &    0.17 &    0.09 & $-$3.26 &  16 &    0.11 &    0.08 & $-$0.12 &   1 &    0.20 &    0.22 &    0.78 &   1 &    0.20 &    0.30 &    0.27 &   1 &    0.30 &    0.31  \\
 BS 16467-062 & $-$3.75 &  52 &    0.15 &    0.04 & $-$3.75 &   3 &    0.16 &    0.14 &    0.48 &   1 &    0.20 &    0.23 & \nodata &   0 & \nodata & \nodata & \nodata &   0 & \nodata & \nodata  \\
\enddata
\end{deluxetable}
\clearpage

\begin{deluxetable}{lrrrrrrrrrrrrrrrrrrrr}
\tablecolumns{21}
\setlength{\tabcolsep}{0.055in}
\tabletypesize{\tiny}
\tablewidth{0pc}
\tablecaption{Abundances relative to Fe (and \ion{Fe}{2} for ionized species) \label{abund2}}
\tablehead{
\colhead{Star}  & \colhead{  [Na/Fe]} & \colhead{numb.} & \colhead{$\sigma$} & \colhead{total}  & \colhead{  [Mg/Fe]} & \colhead{numb.} & \colhead{$\sigma$} & \colhead{total}  & \colhead{  [Al/Fe]} & \colhead{numb.} & \colhead{$\sigma$} & \colhead{total}  & \colhead{  [Si/Fe]} & \colhead{numb.} & \colhead{$\sigma$} & \colhead{total}  & \colhead{  [Ca/Fe]} & \colhead{numb.} & \colhead{$\sigma$} & \colhead{total}
\\
\colhead{ID} & \colhead{} & \colhead{lines} &\colhead{lines} & \colhead{error} & \colhead{} & \colhead{lines} &\colhead{lines} 
& \colhead{error} & \colhead{} & \colhead{lines} &\colhead{lines} & \colhead{error} & \colhead{} & \colhead{lines} &\colhead{lines} 
& \colhead{error} & \colhead{} & \colhead{lines} &\colhead{lines} & \colhead{error}
}
\startdata
   BD+03 740  & \nodata &   0 & \nodata & \nodata &    0.23 &   6 &    0.04 &    0.02 & $-$0.84 &   1 &    0.15 &    0.15 &    0.07 &   1 &    0.15 &    0.15 &    0.38 &  11 &    0.04 &    0.03  \\
  BD+23 3130  & $-$0.09 &   2 &    0.08 &    0.12 &    0.25 &   7 &    0.09 &    0.10 & $-$0.99 &   2 &    0.04 &    0.06 & \nodata &   0 & \nodata & \nodata &    0.24 &  14 &    0.09 &    0.04  \\
  BD+24 1676  & \nodata &   0 & \nodata & \nodata &    0.37 &   6 &    0.07 &    0.03 & \nodata &   0 & \nodata & \nodata &    0.17 &   1 &    0.15 &    0.15 &    0.45 &  14 &    0.08 &    0.03  \\
 BS 16077-007 & \nodata &   0 & \nodata & \nodata &    0.34 &   5 &    0.08 &    0.04 & \nodata &   0 & \nodata & \nodata &    0.13 &   1 &    0.15 &    0.15 &    0.43 &  10 &    0.03 &    0.02  \\
 BS 16080-054 &    0.48 &   2 &    0.15 &    0.12 &    0.44 &   5 &    0.09 &    0.06 & $-$0.70 &   2 &    0.04 &    0.04 &    0.47 &   1 &    0.15 &    0.15 &    0.41 &  11 &    0.06 &    0.05  \\
 BS 16080-093 & \nodata &   0 & \nodata & \nodata &    0.44 &   8 &    0.07 &    0.05 & $-$0.88 &   2 &    0.07 &    0.06 &    0.42 &   1 &    0.15 &    0.15 &    0.27 &   8 &    0.09 &    0.05  \\
 BS 16084-160 & $-$0.22 &   2 &    0.11 &    0.10 &    0.21 &   7 &    0.07 &    0.05 & $-$0.97 &   2 &    0.04 &    0.05 &    0.62 &   1 &    0.15 &    0.15 &    0.18 &  10 &    0.10 &    0.05  \\
 BS 16467-062 & \nodata &   0 & \nodata & \nodata &    0.23 &   4 &    0.10 &    0.11 & $-$0.79 &   2 &    0.04 &    0.06 &    0.32 &   1 &    0.15 &    0.21 &    0.28 &   4 &    0.09 &    0.06  \\
\enddata
\end{deluxetable}
\clearpage

\begin{deluxetable}{lrrrrrrrrrrrrrrrrrrrr}
\tablecolumns{21}
\setlength{\tabcolsep}{0.055in}
\tabletypesize{\tiny}
\tablewidth{0pc}
\tablecaption{Abundances relative to Fe (and \ion{Fe}{2} for ionized species) \label{abund3}}
\tablehead{
\colhead{Star}  & \colhead{[ScII/Fe]} & \colhead{numb.} & \colhead{$\sigma$} & \colhead{total}  & \colhead{  [Ti/Fe]} & \colhead{numb.} & \colhead{$\sigma$} & \colhead{total}  & \colhead{[Ti {\tiny II}/Fe]} & \colhead{numb.} & \colhead{$\sigma$} & \colhead{total}  & \colhead{   [V/Fe]} & \colhead{numb.} & \colhead{$\sigma$} & \colhead{total}  & \colhead{ [V {\tiny II}/Fe]} & \colhead{numb.} & \colhead{$\sigma$} & \colhead{total}
\\
\colhead{ID} & \colhead{} & \colhead{lines} &\colhead{lines} & \colhead{error} & \colhead{} & \colhead{lines} &\colhead{lines} 
& \colhead{error} & \colhead{} & \colhead{lines} &\colhead{lines} & \colhead{error} & \colhead{} & \colhead{lines} &\colhead{lines} 
& \colhead{error} & \colhead{} & \colhead{lines} &\colhead{lines} & \colhead{error}
}
\startdata
   BD+03 740  &    0.29 &   4 &    0.06 &    0.05 &    0.69 &   7 &    0.07 &    0.03 &    0.57 &  28 &    0.11 &    0.06 & \nodata &   0 & \nodata & \nodata &    0.28 &   4 &    0.06 &    0.09  \\
  BD+23 3130  &    0.06 &   6 &    0.06 &    0.05 &    0.21 &  19 &    0.06 &    0.05 &    0.25 &  38 &    0.13 &    0.05 &    0.08 &   3 &    0.25 &    0.16 &    0.12 &   8 &    0.09 &    0.08  \\
  BD+24 1676  &    0.20 &   5 &    0.06 &    0.05 &    0.55 &   9 &    0.08 &    0.03 &    0.45 &  32 &    0.11 &    0.06 & \nodata &   0 & \nodata & \nodata &    0.20 &   5 &    0.09 &    0.09  \\
 BS 16077-007 &    0.31 &   3 &    0.06 &    0.08 &    0.59 &   3 &    0.10 &    0.06 &    0.62 &  23 &    0.09 &    0.07 & \nodata &   0 & \nodata & \nodata &    0.42 &   1 &    0.15 &    0.18  \\
 BS 16080-054 & $-$0.02 &   5 &    0.16 &    0.09 &    0.24 &   6 &    0.10 &    0.07 &    0.30 &  24 &    0.10 &    0.06 & \nodata &   0 & \nodata & \nodata & $-$0.03 &   5 &    0.07 &    0.07  \\
 BS 16080-093 &    0.09 &   3 &    0.06 &    0.07 &    0.38 &   9 &    0.07 &    0.07 &    0.41 &  28 &    0.13 &    0.06 & \nodata &   0 & \nodata & \nodata & $-$0.09 &   1 &    0.15 &    0.17  \\
 BS 16084-160 &    0.09 &   6 &    0.07 &    0.06 & $-$0.02 &  11 &    0.08 &    0.07 &    0.09 &  28 &    0.09 &    0.06 & \nodata &   0 & \nodata & \nodata & $-$0.14 &   6 &    0.06 &    0.07  \\
 BS 16467-062 &    0.31 &   2 &    0.06 &    0.11 &    0.51 &   1 &    0.15 &    0.16 &    0.44 &   9 &    0.12 &    0.10 & \nodata &   0 & \nodata & \nodata & \nodata &   0 & \nodata & \nodata  \\
\enddata
\end{deluxetable}
\clearpage

\begin{deluxetable}{lrrrrrrrrrrrrrrrrrrrr}
\tablecolumns{21}
\setlength{\tabcolsep}{0.055in}
\tabletypesize{\tiny}
\tablewidth{0pc}
\tablecaption{Abundances relative to Fe (and \ion{Fe}{2} for ionized species) \label{abund4}}
\tablehead{
\colhead{Star}  & \colhead{  [Cr/Fe]} & \colhead{numb.} & \colhead{$\sigma$} & \colhead{total}  & \colhead{[Cr {\tiny II}/Fe]} & \colhead{numb.} & \colhead{$\sigma$} & \colhead{total}  & \colhead{  [Mn/Fe]} & \colhead{numb.} & \colhead{$\sigma$} & \colhead{total}  & \colhead{[Mn {\tiny II}/Fe]} & \colhead{numb.} & \colhead{$\sigma$} & \colhead{total}  & \colhead{  [Co/Fe]} & \colhead{numb.} & \colhead{$\sigma$} & \colhead{total}
\\
\colhead{ID} & \colhead{} & \colhead{lines} &\colhead{lines} & \colhead{error} & \colhead{} & \colhead{lines} &\colhead{lines} 
& \colhead{error} & \colhead{} & \colhead{lines} &\colhead{lines} & \colhead{error} & \colhead{} & \colhead{lines} &\colhead{lines} 
& \colhead{error} & \colhead{} & \colhead{lines} &\colhead{lines} & \colhead{error}
}
\startdata
   BD+03 740  & $-$0.07 &   5 &    0.05 &    0.03 &    0.05 &   2 &    0.10 &    0.09 & $-$0.46 &   3 &    0.16 &    0.09 & $-$0.19 &   3 &    0.06 &    0.06 &    0.41 &   3 &    0.09 &    0.05  \\
  BD+23 3130  & $-$0.24 &  19 &    0.10 &    0.04 & $-$0.07 &   4 &    0.16 &    0.10 & $-$0.57 &  10 &    0.13 &    0.05 & $-$0.38 &   3 &    0.07 &    0.06 &    0.20 &   2 &    0.16 &    0.12  \\
  BD+24 1676  & $-$0.13 &   7 &    0.09 &    0.04 & $-$0.09 &   3 &    0.12 &    0.08 & $-$0.43 &   5 &    0.21 &    0.09 & $-$0.30 &   3 &    0.06 &    0.06 &    0.35 &   2 &    0.09 &    0.06  \\
 BS 16077-007 & $-$0.15 &   6 &    0.06 &    0.03 &    0.01 &   2 &    0.10 &    0.11 & $-$0.52 &   3 &    0.16 &    0.09 & $-$0.27 &   2 &    0.06 &    0.09 &    0.34 &   2 &    0.09 &    0.06  \\
 BS 16080-054 & $-$0.43 &   7 &    0.26 &    0.10 & $-$0.21 &   1 &    0.15 &    0.16 & $-$0.61 &   4 &    0.25 &    0.13 & $-$0.59 &   1 &    0.15 &    0.16 &    0.08 &   3 &    0.09 &    0.07  \\
 BS 16080-093 & $-$0.34 &   5 &    0.13 &    0.07 & \nodata &   0 & \nodata & \nodata & $-$0.98 &   3 &    0.16 &    0.10 & \nodata &   0 & \nodata & \nodata &    0.30 &   2 &    0.10 &    0.08  \\
 BS 16084-160 & $-$0.42 &  10 &    0.16 &    0.06 & \nodata &   0 & \nodata & \nodata & $-$0.55 &   9 &    0.25 &    0.09 & $-$0.44 &   3 &    0.06 &    0.07 &    0.09 &   3 &    0.11 &    0.08  \\
 BS 16467-062 & $-$0.46 &   4 &    0.11 &    0.06 & \nodata &   0 & \nodata & \nodata & $-$0.64 &   3 &    0.16 &    0.10 & \nodata &   0 & \nodata & \nodata &    0.69 &   3 &    0.09 &    0.07  \\
\enddata
\end{deluxetable}
\clearpage

\begin{deluxetable}{lrrrrrrrrrrrrrrrrrrrr}
\tablecolumns{21}
\setlength{\tabcolsep}{0.055in}
\tabletypesize{\tiny}
\tablewidth{0pc}
\tablecaption{Abundances relative to Fe (and \ion{Fe}{2} for ionized species) \label{abund5}}
\tablehead{
\colhead{Star}  & \colhead{  [Ni/Fe]} & \colhead{numb.} & \colhead{$\sigma$} & \colhead{total}  & \colhead{  [Cu/Fe]} & \colhead{numb.} & \colhead{$\sigma$} & \colhead{total}  & \colhead{  [Zn/Fe]} & \colhead{numb.} & \colhead{$\sigma$} & \colhead{total}  & \colhead{[Sr {\tiny II}/Fe]} & \colhead{numb.} & \colhead{$\sigma$} & \colhead{total}  & \colhead{ [Y {\tiny II}/Fe]} & \colhead{numb.} & \colhead{$\sigma$} & \colhead{total}
\\
\colhead{ID} & \colhead{} & \colhead{lines} &\colhead{lines} & \colhead{error} & \colhead{} & \colhead{lines} &\colhead{lines} 
& \colhead{error} & \colhead{} & \colhead{lines} &\colhead{lines} & \colhead{error} & \colhead{} & \colhead{lines} &\colhead{lines} 
& \colhead{error} & \colhead{} & \colhead{lines} &\colhead{lines} & \colhead{error}
}
\startdata
   BD+03 740  &    0.06 &   3 &    0.08 &    0.05 & $-$0.80 &   2 &    0.06 &    0.06 & \nodata &   0 & \nodata & \nodata & $-$0.21 &   2 &    0.06 &    0.09 & $-$0.01 &   1 &    0.15 &    0.16  \\
  BD+23 3130  & $-$0.15 &   5 &    0.10 &    0.06 & $-$1.33 &   2 &    0.14 &    0.12 &    0.07 &   2 &    0.04 &    0.08 & $-$0.38 &   2 &    0.10 &    0.08 & $-$0.39 &  11 &    0.08 &    0.05  \\
  BD+24 1676  &    0.03 &   4 &    0.08 &    0.04 & $-$0.83 &   2 &    0.06 &    0.06 &    0.20 &   1 &    0.15 &    0.18 & $-$0.05 &   2 &    0.05 &    0.08 & $-$0.12 &   5 &    0.05 &    0.06  \\
 BS 16077-007 &    0.08 &   3 &    0.08 &    0.05 & $-$0.59 &   1 &    0.15 &    0.16 & \nodata &   0 & \nodata & \nodata &    0.22 &   2 &    0.05 &    0.11 &    0.30 &   1 &    0.15 &    0.17  \\
 BS 16080-054 & $-$0.05 &   3 &    0.11 &    0.07 & \nodata &   0 & \nodata & \nodata & \nodata &   0 & \nodata & \nodata &    0.16 &   2 &    0.05 &    0.17 &    0.12 &   8 &    0.14 &    0.09  \\
 BS 16080-093 &    0.03 &   2 &    0.14 &    0.10 & \nodata &   0 & \nodata & \nodata & \nodata &   0 & \nodata & \nodata & $-$0.13 &   2 &    0.05 &    0.17 & $-$0.28 &   4 &    0.09 &    0.09  \\
 BS 16084-160 & $-$0.13 &   4 &    0.13 &    0.07 & $-$1.64 &   2 &    0.11 &    0.10 & \nodata &   0 & \nodata & \nodata & $-$2.10 &   2 &    0.11 &    0.18 & \nodata &   0 & \nodata & \nodata  \\
 BS 16467-062 &    0.29 &   2 &    0.08 &    0.07 & \nodata &   0 & \nodata & \nodata & \nodata &   0 & \nodata & \nodata & \nodata &   0 & \nodata & \nodata & \nodata &   0 & \nodata & \nodata  \\
\enddata
\end{deluxetable}
\clearpage

\begin{deluxetable}{lrrrrrrrrrrrrrrrr}
\tablecolumns{17}
\setlength{\tabcolsep}{0.055in}
\tabletypesize{\tiny}
\tablewidth{0pc}
\tablecaption{Abundances relative to Fe (and \ion{Fe}{2} for ionized species) \label{abund6}}
\tablehead{
\colhead{Star}  & \colhead{[Zr {\tiny II}/Fe]} & \colhead{numb.} & \colhead{$\sigma$} & \colhead{total}  & \colhead{[Ba {\tiny II}/Fe]} & \colhead{numb.} & \colhead{$\sigma$} & \colhead{total}  & \colhead{[La {\tiny II}/Fe]} & \colhead{numb.} & \colhead{$\sigma$} & \colhead{total}  & \colhead{[Eu {\tiny II}/Fe]} & \colhead{numb.} & \colhead{$\sigma$} & \colhead{total}
\\
\colhead{ID} & \colhead{} & \colhead{lines} &\colhead{lines} & \colhead{error} & \colhead{} & \colhead{lines} &\colhead{lines} 
& \colhead{error} & \colhead{} & \colhead{lines} &\colhead{lines} & \colhead{error} & \colhead{} & \colhead{lines} &\colhead{lines} 
& \colhead{error} 
}
\startdata
   BD+03 740  &   0.36 &   2 &   0.11 &   0.10 &$-$0.37 &   1 &   0.15 &   0.17 &\nodata &   0 &\nodata &\nodata &$<$0.68 &   1 &\nodata &\nodata  \\
  BD+23 3130  &   0.00 &  10 &   0.12 &   0.06 &$-$0.47 &   3 &   0.06 &   0.09 &$-$0.26 &   1 &   0.15 &   0.17 &   0.31 &   1 &   0.10 &   0.11  \\
  BD+24 1676  &   0.31 &   5 &   0.14 &   0.09 &$-$0.33 &   2 &   0.06 &   0.09 &\nodata &   0 &\nodata &\nodata &$<$0.59 &   1 &\nodata &\nodata  \\
 BS 16077-007 &   0.85 &   2 &   0.19 &   0.16 &$-$0.11 &   1 &   0.15 &   0.18 &\nodata &   0 &\nodata &\nodata &$<$0.81 &   1 &\nodata &\nodata  \\
 BS 16080-054 &   0.31 &   9 &   0.19 &   0.10 &$-$0.42 &   4 &   0.06 &   0.08 &\nodata &   0 &\nodata &\nodata &$<$0.23 &   1 &\nodata &\nodata  \\
 BS 16080-093 &   0.17 &   1 &   0.15 &   0.17 &$-$0.27 &   2 &   0.06 &   0.09 &\nodata &   0 &\nodata &\nodata &   0.20 &   1 &   0.20 &   0.21  \\
 BS 16084-160 &\nodata &   0 &\nodata &\nodata &$-$1.99 &   2 &   0.09 &   0.10 &\nodata &   0 &\nodata &\nodata &$<-$0.25 &   1 &\nodata &\nodata  \\
 BS 16467-062 &\nodata &   0 &\nodata &\nodata &$<-$0.68 &   1 &\nodata &\nodata &\nodata &   0 &\nodata &\nodata &$<$1.24 &   1 &\nodata &\nodata  \\
\enddata
\end{deluxetable}
\clearpage
\end{landscape}

\begin{deluxetable}{llrrc}
\tablecolumns{5}
\tabletypesize{\scriptsize}
\tablewidth{0pc}
\tablecaption{Neutron-capture abundances of CS 31078-018 \label{ncapt}}
\tablehead{
       \colhead{Atomic}  &  \colhead{Element} &  \colhead{[X/Fe]} &
       \colhead{log$\epsilon$(X)} & \colhead{error}\\ \colhead{Number} &
       \colhead{Name} & \colhead{} & \colhead{} & \colhead{} }
\startdata    
38  &  Sr {\scriptsize II} &    0.31  &    0.30 & 0.18\\
39  &  Y {\scriptsize II}  &    0.23  & $-$0.43 & 0.15\\
40  &  Zr {\scriptsize II} &    0.62  &    0.32 & 0.14\\
44  &  Ru {\scriptsize I}  &    0.65  & $-$0.35 & 0.19\\
46  &  Pd {\scriptsize I}  &    0.79  & $-$0.36 & 0.19\\
56  &  Ba {\scriptsize II} &    0.72  & $-$0.06 & 0.31\\
57  &  La {\scriptsize II} &    0.74  & $-$1.00 & 0.22\\
58  &  Ce {\scriptsize II} &    0.66  & $-$0.67 & 0.21\\
60  &  Nd {\scriptsize II} &    0.79  & $-$0.62 & 0.21\\
62  &  Sm {\scriptsize II} &    1.13  & $-$0.77 & 0.16\\
63  &  Eu {\scriptsize II} &    1.23  & $-$1.17 & 0.17\\
64  &  Gd {\scriptsize II} &    1.03  & $-$0.76 & 0.17\\
66  &  Dy {\scriptsize II} &    1.00  & $-$0.77 & 0.15\\
67  &  Ho {\scriptsize II} &    1.54  & $-$1.11 & 0.22\\
68  &  Er {\scriptsize II} &    1.02  & $-$0.96 & 0.20\\
70  &  Yb {\scriptsize II} &    0.76  & $-$1.07 & 0.21\\
90  &  Th {\scriptsize II} &    1.47  & $-$1.35 & 0.25
\enddata
\end{deluxetable}

\begin{deluxetable}{lrrrl}
\tablecolumns{5}
\tabletypesize{\scriptsize}
\tablewidth{0pc}
\tablecaption{A sample of highly neutron-capture deficient stars \label{ncaptlow}}
\tablehead{
       \colhead{Star}  &  \colhead{[Fe/H]} &  \colhead{[Sr/H]} &
       \colhead{[Ba/H]} & \colhead{source}\\ \colhead{ID} &
       \colhead{} & \colhead{} & \colhead{} & \colhead{} }
\startdata    
BS 16084-160 &$-$3.26\tablenotemark{*}  &  $-$5.29 & $-$5.25 &  1  \\
CS 30336-049 &$-$4.16\tablenotemark{*}  &  $-$5.59 & $-$5.48 &  1   \\
CS 29502-092 &$-$3.20\tablenotemark{*}  &  $-$3.59 & $-$4.46 &  1   \\
HD 4306      &$-$2.89  &  $-$3.00 & $-$4.06 &  2 \\
BS 16469-075 &$-$3.03  &  $-$2.80 & $-$4.15 &  2	    \\
BS 16920-017 &$-$3.12  &  $-$3.55 & $-$4.95 &  2	    \\
BS 16928-053 &$-$2.91  &  $-$3.14 & $-$4.07 &  2	    \\
Draco 119    &$-$2.97  & $<-$5.47 & $<-$5.57 &  3  \\
CS 22169-035 &$-$3.04  &  $-$3.12 & $-$4.23 &  4	    \\
CS 22172-002 &$-$3.86  &  $-$5.17 & $-$5.03 &  4	    \\
CS 22189-009 &$-$3.49  &  $-$4.44 & $-$4.78 &  4	    \\
CS 22897-008 &$-$3.41  &  $-$2.97 & $-$4.41 &  4	    \\
CS 22952-015 &$-$3.43  &  $-$4.42 & $-$4.76 &  4	    \\
CS 22968-014 &$-$3.56  &  $-$5.36 & $-$5.33 &  4	    \\
CS 29502-042 &$-$3.19  &  $-$5.17 & $-$4.88 &  4	    \\
CS 30325-094 &$-$3.30  &  $-$5.54 & $-$5.21 &  4     \\
BS 16934-002 &$-$2.81\tablenotemark{*}  &  $-$4.13 & $-$4.61 &  5  \\
CS 30327-038 &$-$2.64\tablenotemark{*}  &  $-$3.28 & $-$4.14 &  5  \\     
HE 1356-0622 &$-$3.36\tablenotemark{*}  &  $-$5.24 & $-$4.45 &  5  
\enddata
\tablenotetext{*}{[Fe II/H]}
\tablerefs{(1) This study; (2) \citet{honda04}; (3)
  \citet{fulbright04}; (4) \citet{francois}; (5) \citet{aoki05}; (6) \citet{cohen07-2}}
\end{deluxetable}

\newpage

\clearpage

\clearpage

\clearpage

\end{document}